\documentclass[aps,preprint,showpacs,preprintnumbers,amsmath,amssymb]{revtex4}
\usepackage{graphicx}
\usepackage{hyperref}
 
\def\beq{\begin{equation}}
\def\eeq{\end{equation}}
\def\beqn{\begin{eqnarray}}
\def\eeqn{\end{eqnarray}}

\def\bseq{\begin{subequations}}
\def\eseq{\end{subequations}}

\def\1 {{\bf 1}}
\def\0 {{\bf 0}}
\def\3 {{\bf 3}}

\def\r {{\bf r}}

\def\n {{\bf n}}
\def\u {{\bf u}}
\def\v {{\bf v}}
\def\m {{\bf m}}

\def\R {{\bf R}}
\def\L {{\bf L}}
\def\C {{\bf C}}
\def\H {{\bf H}}

\def\O {{\bf O}}

\def\R {{\bf R}}

\def\bcalH {\mbox{\boldmath $\mathcal H$}}
\def\bcalL {\mbox{\boldmath $\mathcal L$}}
\def\bcalR {\mbox{\boldmath $\mathcal R$}}
\def\bcalP {\mbox{\boldmath $\mathcal P$}}
\def\bcalM {\mbox{\boldmath $\mathcal M$}}
\def\brho {\mbox{\boldmath $\rho$}}
\def\bphi {\mbox{\boldmath $\phi$}}

\def\bmu {\mbox{\boldmath $\mu$}}
\def\bOmega {\mbox{\boldmath $\Omega$}}
\def\bkappa {\mbox{\boldmath $\kappa$}}
\def\bSig {\mbox{\boldmath $\Sigma$}}

\begin{document}

\title{Unified view on linear response of 
interacting identical and distinguishable
particles from
multiconfigurational time-dependent Hartree methods}

\author{Ofir E. Alon$^{1}$\footnote{E-mail: ofir@research.haifa.ac.il},
Alexej I. Streltsov$^{2}$\footnote{E-mail: Alexej.Streltsov@pci.uni-heidelberg.de},
and Lorenz S. Cederbaum$^{2}$\footnote{E-mail: Lorenz.Cederbaum@pci.uni-heidelberg.de}}

\affiliation{$^{1}$Department of Physics, University of Haifa at Oranim, Tivon
36006, Israel}

\affiliation{$^{2}$Theoretische Chemie, Physikalisch-Chemisches Institut, Universit\"at
Heidelberg, D-69120 Heidelberg, Germany}

\date{\today}

\begin{abstract}
A unified view on linear response of interacting systems 
utilizing multicongurational time-dependent
Hartree methods is presented. 
The cases of one-particle and two-particle response
operators for identical 
particles and up to all-system response operators for distinguishable
degrees-of-freedom are considered. 
The working equations for systems of identical bosons (LR-MCTDHB) 
and identical fermions (LR-MCTDHF), 
as well for systems of distinguishable particles (LR-MCTDH) are explicitly derived. 
These linear-response theories provide 
numerically-exact
excitation energies and system's properties, 
when numerical convergence is achieved in the calculations.
\end{abstract}

\maketitle

%%%%%%%%%%%%%%%%%%%%%%%%%%%%%%%%%%%%%%%%%%%%%%%%
\section{Introduction}\label{intro_sec}
%%%%%%%%%%%%%%%%%%%%%%%%%%%%%%%%%%%%%%%%%%%%%%%%

Excitation spectra of quantum systems are one of the widely used notions in
our modern quantum world. 
They often allow one to explore the microscopic structure
of various quantum objects, 
such as nuclei, atoms, molecules, and solids, and more recently,
of ultracold quantum gases and Bose-Einstein condensates 
\cite{book_exc_1,book_exc_2,book_exc_3,book_exc_4,book_exc_5,book_exc_6,book_exc_7,book_exc_8}. 
The excitation spectrum and excited states govern the spectroscopy and
thermal properties of the quantum system of interest,
and, of course, its quantum dynamics.

It is generally not possible to compute the excitation spectrum of the quantum 
object under interest analytically,
except for a select few systems and models. 
One is opt then to resort to theoretical many-particle methods and to their numerical implementation. 
An often appealing and practical way to compute the excitation spectrum is linear response.
Linear response is an in-principle exact approach, see, e.g., Refs.~\cite{Fetter_book,LR_JCP2},
namely that one can obtain the exact excitation spectrum from the exact ground state,
when weak perturbing fields are applied.

Notably the most famous linear-response theory is
that for interacting identical fermions (electrons) 
whose wavefunction is described (approximated) by a single determinant;
it is termed in different formulations 
time-dependent Hartree-Fock or 
random-phase approximation \cite{HF_LR_Dirac,Ripka_book,Ring_book}.
In many cases the ground state of electronic systems
cannot be adequately described by a single determinant (configuration),
and thus a multiconfigurational description becomes necessary --
even more so for the linear-response computed excitations.
In the context of atomic and molecular physics,
linear response is often used in conjunction with
self-consistent multiconfigurational and configuration-interaction ans\"atze,
see, e.g., Refs.~\cite{LR_JCP1,LR_JCP2,LR_JCP4}, 
and the review article Ref.~\cite{Olsen_Review} 
and references therein.
We mention that for molecular systems 
and distinguishable degrees-of-freedom (vibrations),
linear response for multiconfigurational 
wavefunctions has been 
put forward, 
see Refs.~\cite{LR_JCP3,LR_JCP5}.

For interacting identical bosons,
which are traditionally assumed to be well described by a single permanent, mean-field wavefunction with all the bosons 
occupying the one and the same orbital,
linear response also goes way back and
is termed in different formulations 
Bogoliubov approximation or 
Bogoliubov--de Gennes equations \cite{BdG,Book_Pitaevskii,Book_Pethick}.
With the advent of Bose-Einstein condensates made of ultracold quantum gases,
this standard linear response has drawn much attention 
\cite{LR_GP_Ruprecht,Esry,Castin_Dum,Book_Pitaevskii,Book_Pethick}.

It turns out that, unlike fermions, there are many distinct mean fields for bosons, 
because any number of bosons can occupy one, two, or many different orbitals.
In other words, whereas a single configuration for fermions
(i.e., a determinant, where each fermion ought to occupy a different orbital) is the only mean field for fermions, 
there are many distinct single configurations (i.e., permanents) for bosons, 
and hence many bosonic mean fields. 
For this plurality of bosonic mean fields
and how to find among them the best (i.e., energetically lowest) 
mean field, see Refs.~\cite{BMF,Pathway}.
In the context of linear response for bosonic systems,
this has recently led to the linear-response
theory of the best mean field \cite{LR_BMF},
which unraveled excitations that cannot be found
within the standard linear-response theory 
for bosons.

Before we proceed further,
it should be mentioned for completeness that there is a huge 
literature on the computation and exploration
of excitation spectra and excited states
based on a variety of methods
(either direct or/and in combination with linear response) such as:
Configuration Interaction (also sometimes referred to as exact diagonalization);
Coupled Cluster;
Green's Functions;
Density Functional Theory;
Quantum Monte Carlo; 
and more.
These are not the direct subject of the present work.

In the present work we develop linear-response theories by linearizing
numerically-exact evolution methods to solve 
the time-dependent
many-particle Schr\"odinger equation.
To be concrete, we derive and present a unified view of linear-response theories
from the point of view of multiconfigurational time-dependent Hartree (MCTDH) propagation methods.
The MCTDH method was invented in Refs.~\cite{MCTDH_cpl,MCTDH_jcp} and is
widely used for multi-dimensional dynamical systems
consisting of distinguishable degrees-of-freedom, typically vibrations,
see, e.g., Refs.~\cite{MCTDH_app1,MCTDH_app2,MCTDH_app3,MCTDH_app4,MCTDH_app5,
MCTDH_app6,MCTDH_app7,MCTDH_app8},
the reviews Refs.~\cite{MCTDH_review,MCTDH_book} and references therein, 
and the software package Ref.~\cite{MCTDH_Package}.
MCTDH is considered at present the most efficient 
approach for accurate wavepacket
propagation for distinguishable particles.
MCTDH has been successfully extended to treat identical particles by accounting explicitly 
for the quantum statistics of the particles. 
MCTDH for fermions (MCTDHF) was derived in Refs.~\cite{MCTDHF1,MCTDHF2,MCTDHF3,MCTDHF4,unified};
see, e.g., Refs.~\cite{MCTDHF_app1,MCTDHF_app2,MCTDHF_app3,
MCTDHF_app4,MCTDHF_app5,MCTDHF_app6,MCTDHF_app7,MCTDHF_app8,MCTDHF_app9,MCTDHF_app10} for applications.
MCTDH for bosons (MCTDHB) 
was derived in Refs.~\cite{MCTDHB1,MCTDHB2};
for applications see, e.g., Refs.~\cite{MCTDHB_app1,MCTDHB_app2,MCTDHB_app3,MCTDHB_app4,
MCTDHB_app5,MCTDHB_app6}, 
and the software package Ref.~\cite{MCTDHB_Package}.
We would like to mention that
MCTDH itself has
recently been applied with much success to 
a few-particle bosonic and fermionic systems,
see, e.g., Refs.~\cite{MCTDH_id1,MCTDH_id2,MCTDH_id3,MCTDH_id4,MCTDH_id5,MCTDH_id6}.

The equations of motion in MCTDH propagation methods originate directly
from the time-dependent (Dirac-Frenkel) variational principle. 
As such, the coupled
time-dependent orbitals and expansion coefficients,
and hence the many-particle wavefunction itself, 
are propagated in time in an optimal manner.
Prescribing the linear response 
of the MCTDH propagation methods is hence a natural task.
The MCTDH theories are very efficient, well-defined, commonly used, 
and formally exact,
which also motivates us here to present the fundamental theory
of their linearization.

Most recently, we have undertaken the respective task for bosons and presented the
linear-response theory of MCTDHB \cite{LR_MCTDHB}.
It is a self-consistent multiconfigurational linear-response theory 
capable of computing exact many-body excitations of identical bosons,
thus generalizing and extending the amply used, standard linear-response theory 
for Bose-Einstein condensates -- 
Bogoliubov--de Gennes equations \cite{BdG,Book_Pitaevskii,Book_Pethick}.
With increased capabilities to accurately measure excitation spectra of Bose-Einstein
condensates in various trap potentials,
see in this context Refs.~\cite{EXBEC1,EXBEC2,EXBEC3,EXBEC4},
much is to be anticipated from the multiconfigurational linear-response theory of MCTDHB.
The purpose of the present work 
is to build atop and expand on our recent findings for bosons.
In what follows we derive the respective linear-response theories
of the MCTDH propagation methods.
This is done by linearizing the corresponding numerically-exact equations of motion
for identical particles and distinguishable degrees-of-freedom,
in a unified representation and manner.

The structure of the paper is as follows.
In Sec.~\ref{SE_LR} we present a deductive exposition 
of the general feature that linear response atop the exact ground state 
provides the exact excitation spectra. 
We do so by treating the (already linear)
time-dependent many-particle Schr\"odinger equation.
In Sec.~\ref{indis_sec} we present the core linear-response theory
for identical particles (bosons and fermions) based on MCTDHB and MCTDHF.
We treat explicitly in the various subsections the different ingredients,
up to the matrix form of the linear-response equations and
utilizing its properties to solve the time-dependent
identical-particle Schr\"odinger equation in linear response (Subsec.~\ref{matrix_form_sec}).
We also utilize the projector operators in these MCTDH theories
to arrive at an orthogonal response space,
in which the orbitals' and coefficients' response amplitudes 
are explicitly orthogonal to the ground-state wavefunction.
In Sec.~\ref{dis_sec} we develop the linear-response
theory for distinguishable degrees-of-freedom
based on MCTDH,
and discuss its structure and relation to the identical-particle theories.
As a complementary result of the linear-response theories of
Secs.~\ref{indis_sec} and \ref{dis_sec}, 
we introduce into the MCTDH methods 
the notion of ``fully-projected'' equations of motion,
which are obtained by adding to the famous orbitals' differential condition 
\cite{MCTDH_cpl,MCTDH_jcp} a complementary differential 
condition onto the coefficients' part.  
We summarize and conclude in Sec.~\ref{sum_con_sec}.
Finally, the two appendixes present 
some mathematical supplementaries. 
Appendix \ref{dif_con_proof_appendix} deals with the (orbital) differential condition in 
the MCTDHB and MCTDHF theories, 
and Appendix \ref{MCTDH_tensor_appendix} with the tensor-product representation of quantities in 
the MCTDH theory.

%%%%%%%%%%%%%%%%%%%%%%%%%%%%%%%%%%%%%%%%%%%%%%%%
\section{Linear Response of the time-dependent 
(many-particle) Schr\"odinger Equation}\label{SE_LR}
%%%%%%%%%%%%%%%%%%%%%%%%%%%%%%%%%%%%%%%%%%%%%%%%

We wish to show in this section, 
as a deductive and complementary preamble, 
that the linear response
of the time-dependent Schr\"odinger equation with respect to perturbation of, 
typically, the ground state
gives rise to the {\it exact excitation spectrum} 
and corresponding eigenfunctions of the quantum system.
Actually, no assumptions are made on the quantum system's Hamiltonian, 
except that the system is perturbed by a weak time-periodic field, hence -- linear response.
The derivation is inspired by \cite{LR_GP_Ruprecht,Castin_Dum,LR_BMF}.
An additional purpose of this section is to introduce and clarify in a simpler, linear problem
the role of projection operator(s) leading to fully orthogonal,
or ``fully projected'' as we shall refer to it below, 
dynamics.
This will become instrumental 
later on.

Let the time-independent Hamiltonian and Schr\"odinger equation read:
\beq\label{SE_1}
 \hat H \Phi_k = E_k \Phi_k, \qquad k=0,1,2,\dots,
\eeq
with the eigenvalues $E_k$ and eigenvectors $\Phi_k$ and for $k=0$ (typically) the ground state.
We may write Eq.~(\ref{SE_1}) 
equivalently as $\hat P_{\Phi_k}\hat H \Phi_k =0, k=0,1,2,\dots$ with
the projector $\hat P_{\Phi_k} = 1 - |\Phi_k\rangle\langle\Phi_k|$, also see below.

Now we wish 
to solve the time-dependent Schr\"odinger equation with the
weak
time-dependent ($\omega>0$) perturbation
\beq\label{SE_2}
 \hat H(t)\Psi(t) = i\dot\Psi(t), \qquad  \hat H(t) = \hat H + \hat f^\dag e^{-i \omega t} + \hat f e^{+i \omega t}.
\eeq
We first perform a unitary transformation.
By making the following 
assignment:
\beq\label{SE_3}
 \Psi(t) \Longrightarrow \Psi(t) e^{-i \int^t \langle\Psi(t')|\hat H(t')|\Psi(t')\rangle dt'}
\eeq
we obtain the ``projected" time-dependent Schr\"odinger equation
\beq\label{SE_4}
 \hat P_{\Psi} \hat H(t)\Psi(t) = i\dot\Psi(t), \qquad  \hat P_{\Psi} = 1 - |\Psi(t)\rangle\langle\Psi(t)|.
\eeq
The ``projected" Schr\"odinger equation, Eq.~(\ref{SE_4}), 
has the appealing property 
that the time-evolution is completely orthogonal in the sense that:
\beq\label{SE_5}
  i\langle\Psi(t)|\dot\Psi(t)\rangle = 0.
\eeq
The differential condition Eq.~(\ref{SE_5}) 
would relate later on to 
the familiar orbital differential condition introduced in Refs.~\cite{MCTDH_cpl,MCTDH_jcp}
and the below introduced coefficients' 
differential condition in MCTDHB and MCTDHF, 
and in MCTDH theories.

In linear response we write for the solution of 
the time-dependent Schr\"odinger equation (\ref{SE_4})
[or, simply of Eq.~(\ref{SE_2}) but then with the additional global phase $e^{-i E_0 t}$] 
the ansatz:
\beq\label{SE_6_LR_ansatz}
  \Psi(t) \approx \Phi_0 + Ue^{-i \omega t} + V^\ast e^{+i \omega t}.
\eeq
Here and hereafter, the stationary state $\Phi_0$ is the zeroth-order approximation to
$\Psi(t)$ and the $U$ and $V^\ast$ are the first-order corrections or perturbations which
are assumed to be small.
Accordingly, in what follows the equations to be derived will be referred to as zeroth-order equations
if they include only zeroth-order terms.
The linear-response equations as well as other properties to be discussed will be referred 
to as first-order equations,
since they are linear in the perturbations.

Making use of the differential condition Eq.~(\ref{SE_5}), 
we immediately arrive
at the orthogonality of the perturbed parts (for $\omega>0$) 
of the time-dependent wavefunction Eq.~(\ref{SE_6_LR_ansatz}) with
respect to the ground state:
\beq\label{SE_7}
 \langle\Phi_0|U\rangle = 0, \quad \ \langle\Phi_0|V^\ast\rangle = 0  \qquad \Longleftrightarrow \qquad
 \hat P_{\Phi_0} U  = U, \quad \   \hat P_{\Phi_0} V^\ast = V^\ast.
\eeq

Inserting the ansatz Eq.~(\ref{SE_6_LR_ansatz}) into Eq.~(\ref{SE_4}),
we obtain to zeroth order the stationary problem for the ground state,
$\hat P_{\Phi_0}\hat H \Phi_0 = 0$.
Furthermore,
making use of Eq.~(\ref{SE_7})
we arrive at the following result to first order:
\beq\label{SE_inter_t_exps}
  \hat P_{\Phi_0}  (\hat H - E_0) (Ue^{-i \omega t} + V^\ast e^{+i \omega t}) 
 + \hat P_{\Phi_0} (\hat f^\dag e^{-i \omega t} + \hat f e^{+i \omega t})|\Phi_0\rangle 
= \omega (Ue^{-i \omega t} - V^\ast e^{+i \omega t}).
\eeq 
Next, equating same powers of $e^{\mp i \omega t}$,
adding a (redundant) projector in front of the perturbed 
parts of the wavefunction, $U$ and $V^\ast$,
and collecting in a matrix form,
we get as the final result for the equation 
of the perturbed time-dependent 
wavefunction:
\beq\label{SE_8_LR_EOM}
\left[\begin{pmatrix} 
\hat P_{\Phi_0} (\hat H - E_0) \hat P_{\Phi_0} & 0 \\
0 &  -  \hat P_{\Phi_0}^\ast (\hat H^\ast -E_0) \hat P_{\Phi_0}^\ast \\
\end{pmatrix} - \omega 
\begin{pmatrix} 
1 & 0 \\
0 & 1 \\
\end{pmatrix}
\right]
\begin{pmatrix} 
U \\
V \\
\end{pmatrix} =
\begin{pmatrix} 
- \hat P_{\Phi_0} \hat f^\dag |\Phi_0\rangle \\
 \hat P_{\Phi_0}^\ast \hat f^\ast |\Phi_0^\ast\rangle \\
\end{pmatrix}.
\eeq

To solve for Eq.~(\ref{SE_8_LR_EOM}) we first introduce the linear-response matrix $\bcalL$:
\beq\label{L_SE_intro} 
\bcalL =  \begin{pmatrix} 
\hat P_{\Phi_0} (\hat H - E_0) \hat P_{\Phi_0} & 0 \\
0 &  -  \hat P_{\Phi_0}^\ast (\hat H^\ast -E_0) \hat P_{\Phi_0}^\ast \\
\end{pmatrix} 
\eeq
and solve the linear-response eigenvalue system:
\beq\label{SE_9_LR_FROM}
\bcalL
\begin{pmatrix} 
U_k \\
V_k \\
\end{pmatrix}
= \omega_k 
\begin{pmatrix} 
U_k \\
V_k \\
\end{pmatrix}
\eeq
associated with the Schr\"odinger equation.
It is easily diagonalized and has two ``branches":
\beqn\label{SE_10_LR_Eigen}
& & 
\begin{pmatrix} 
U_k \\
V_k \\
\end{pmatrix} =
\begin{pmatrix} 
|\Phi_k\rangle \\
 0 \\
\end{pmatrix}, 
\qquad \omega_k = E_k - E_0, \qquad k=0,1,2,\ldots, \nonumber \\
& &
\begin{pmatrix} 
U_{-k} \\
V_{-k} \\
\end{pmatrix} =
\begin{pmatrix} 
 0 \\
 |\Phi_{k}^\ast\rangle \\
\end{pmatrix}, 
\qquad - \omega_k, \qquad k=0,1,2,\ldots. \
\eeqn
Thus, it is readily seen that the linear-response of the Schr\"odinger equation (\ref{SE_2})
provides the exact excitation energies and corresponding excited-state eigenfunctions 
with respect to $\Phi_0$,
which is usually the ground state.
The so-called zero-mode excitations, 
which equal nothing but the 
ground-state energy of the system,
separate the positive [$w_k>0$; upper ``branch" of Eq.~(\ref{SE_10_LR_Eigen})] 
and negative [$w_k<0$; lower ``branch" of Eq.~(\ref{SE_10_LR_Eigen})]
parts of the spectrum.

Having solved the linear-response eigenvalue system, Eq.~(\ref{SE_9_LR_FROM}), 
we can now express the solution of Eq.~(\ref{SE_8_LR_EOM}) 
in terms of the excitation energies $\{\omega_k\}$ and eigenvectors 
$\left\{\begin{pmatrix} 
U_{k} \\
V_{k} \\
\end{pmatrix}
\right\}$.
To this end we expand the perturbed wavefunction:
\beq\label{SE_11_wavefunction}
\begin{pmatrix} 
U \\
V \\
\end{pmatrix} = \sum_k c_k
\begin{pmatrix} 
U_k \\
V_k \\
\end{pmatrix} = 
\sum_{k>0} 
\left\{
c_k
\begin{pmatrix} 
|\Phi_k\rangle \\
 0 \\
\end{pmatrix} +
c_{-k}
\begin{pmatrix} 
0 \\
|\Phi_k^\ast\rangle \\
\end{pmatrix}
\right\}.
\eeq
The zero-mode, two $k=0$ excitations do not have to be included in the expansion Eq.~(\ref{SE_11_wavefunction}),
because they give no contribution in view of the orthogonality relations in Eq.~(\ref{SE_7}).
The expansion coefficients thus read:
\beqn\label{SE_12_LR_expansion}
& & c_k = \frac{\langle\Phi_k|\hat f^\dag|\Phi_0\rangle}{w-w_k}, \qquad k=1,2,\ldots, \nonumber \\
& & c_{-k} = - \frac{\langle\Phi_{k}^\ast|\hat f^\ast|\Phi_0^\ast\rangle}{w+w_{k}}, \qquad k=1,2,\ldots, \
\eeqn
which completes
the solution of the Schr\"odinger equation within 
linear response.

A final remark concerning hermiticity and completeness is in place.
The linear-response of the (already linear) 
Schr\"odinger equation gives rise to
the eigenvalue system Eq.~(\ref{SE_9_LR_FROM}) with an hermitian linear-response matrix 
$\bcalL$, Eq.~(\ref{L_SE_intro}) (for other characteristics of $\bcalL$ see above). 
Thus, the usual properties of normalization, orthogonality, and completeness of its 
eigenfunctions $\left\{\begin{pmatrix} U_k \\ V_k \\ \end{pmatrix}\right\}$ simply hold.
From them, the resolution of the identity:
\beq\label{SE_12_LR_identity}
\1 = 
\begin{pmatrix} 
|\Phi_0\rangle \\
 0\\
\end{pmatrix}
\begin{pmatrix} 
\langle\Phi_0| \ 0 \\
\end{pmatrix} +
\begin{pmatrix} 
0 \\
|\Phi_0^\ast\rangle \\
\end{pmatrix}
\begin{pmatrix} 
 0 \ \langle\Phi_0^\ast| \\
\end{pmatrix}
+ \sum_{k>0} \left\{
\begin{pmatrix} 
|\Phi_k\rangle \\
 0\\
\end{pmatrix}
\begin{pmatrix} 
\langle\Phi_k| \ 0 \\
\end{pmatrix} +
\begin{pmatrix} 
0 \\
|\Phi_k^\ast\rangle \\
\end{pmatrix}
\begin{pmatrix} 
 0 \ \langle\Phi_k^\ast| \\
\end{pmatrix} \right\}
\eeq
and spectral resolution of the linear-response matrix:
\beq\label{SE_13_LR_L_matrix}
\bcalL = 
\sum_{k>0} \omega_k \left\{
  \begin{pmatrix} 
|\Phi_k\rangle \\
 0\\
\end{pmatrix}
\begin{pmatrix} 
\langle\Phi_k| \ 0 \\
\end{pmatrix} -
\begin{pmatrix} 
0 \\
|\Phi_k^\ast\rangle \\
\end{pmatrix}
\begin{pmatrix} 
 0 \ \langle\Phi_k^\ast| \\
\end{pmatrix} \right\}
\eeq
follow.
In the forthcoming sections, 
we discuss extensively the linear response of many-particle
systems described by multiconfigurational time-dependent Hartree methods,
which of course consist of nonlinear equations.
The resulting response matrices, as in, e.g., 
Refs.~\cite{LR_GP_Ruprecht,Castin_Dum,LR_BMF}, 
will no longer be hermitian.
Hence, analogous resolutions like 
Eqs.~(\ref{SE_12_LR_identity}) and (\ref{SE_13_LR_L_matrix}) 
would be assumed to hold and then constructed.

%%%%%%%%%%%%%%%%%%%%%%%%%%%%%%%%%%%%%%%%%%%%%%%%
%%%% \section{LR-MCTDHB and LR-MCTDHF: LR-MCTDHX}\label{indis_sec}
\section{Linear response in the multiconfigurational time-dependent Hartree framework for identical particles}\label{indis_sec}
%%%%%%%%%%%%%%%%%%%%%%%%%%%%%%%%%%%%%%%%%%%%%%%%

This section deals with systems of interacting bosons or fermions and --
starting from the propagation theories MCTDHB and MCTDHF --
it derives their respective linear-response theories, 
which we denote by LR-MCTDHB and LR-MCTDHF,
in a unified framework.

%%%%%%%%%%%%%%%%%%%%%%%%%%%%%%%%%%%%%%%%%%%%%%%%
\subsection{Basic and new ingredients}\label{Basic_n_ingr_sec}
%%%%%%%%%%%%%%%%%%%%%%%%%%%%%%%%%%%%%%%%%%%%%%%%

Consider a quantum object made of $N$ identical interacting particles, bosons or fermions.
Our starting point is the MCTDHX (X=B,F) equations of motion, 
see Refs.~\cite{MCTDHB1,MCTDHB2,MCTDHF1,MCTDHF2,MCTDHF3,MCTDHF4},
which can be represented in a unified manner \cite{unified}.
The MCTDHX wavefunction is written as $|\Psi(t)\rangle = \sum_{\vec n} C_{\vec n} |\vec n;t\rangle$,
where $|\vec n;t\rangle = \frac{1}{\sqrt{n_1! \ldots n_M!}} \left[\hat c_1(t)^\dag\right]^{n_1}
\cdots \left[\hat c_M(t)^\dag\right]^{n_M}|vac\rangle$ are time-dependent configurations -- 
either permanents for bosons or determinants for fermions.
$\vec n = (n_1,\ldots,n_M)$ is the vector of occupations and $|vac\rangle$ the vacuum state.
The creation $\{\hat c_k^\dag(t)\}$ and respective annihilation operators $\{\hat c_k(t)\}$
are associated with a set of $M$ time-dependent orthonormal orbitals $\{\phi_k(\r,t)\}$. 
The vector of time-dependent coefficients $\C(t)=\{C_{\vec n}\}$ runs over all time-dependent configurations
generated by distributing the $N$ identical particles over the $M$ orbitals.
The number of such configurations is denoted hereafter by $N_{\mathit{conf}}$.

The MCTDHX equations of motion are given by:
\beqn\label{MCTDHX_equ}
& &  \hat {\mathbf P} \sum_{q=1}^M [\rho_{kq} \hat h +
 \sum^M_{s,l=1}\rho_{kslq} \hat{W}_{sl}] |\phi_q\rangle = i \sum_{q=1}^M \rho_{kq} |\dot\phi_q\rangle, \qquad k=1,\ldots,M, \nonumber \\
& & {\mathbf H}(t)\C(t) = i\frac{\partial \C(t)}{\partial t} \qquad
\Longleftrightarrow \qquad
\C^{\hat H}(t) = i \dot \C(t), \qquad
 H_{\vec{n}\vec{n}'}(t) = \langle\vec{n};t|\hat H|\vec{n}';t\rangle, \
\eeqn
with the local (direct) potentials:
\beq\label{local}
 \hat{W}_{kq} = \int d\r' \phi^\ast_k(\r') \hat W(\r-\r') \phi_q(\r').
\eeq
Both reduced density matrices \cite{Lowdin,Yukalov_book,Mazziotti_book},
which are highly helpful in MCTDHX theories \cite{MCHB}, 
and mapping of configurations and the operation of operators upon them \cite{Mapping,3well}
are used to shorten and simplify the notation.
Occasionally we will use a double or mixed notation,
especially for the equations of motion of the coefficients, 
when instructive.
The first notation implies, throughout this work:
$\rho_{kq}=\langle \Psi|\hat \rho_{kq}|\Psi\rangle$ and 
$\rho_{kslq}=\langle \Psi|\hat \rho_{kslq}|\Psi\rangle$,
where
$\hat \rho_{kq}=\hat c_k^\dag \hat c_q$ and
$\hat \rho_{kslq} = \hat c_k^\dag \hat c_s^\dag \hat c_l \hat c_q$
are the density operators. 
The second notation means, throughout this work, as follows.
Consider a generic, second-quantized operator $\hat O$ acting on the MCTDHX wavefunction.
The operation readdresses the configurations,
multiples them by respective matrix elements and numerical factors,
and thereby changes the
vector of coefficients \cite{Mapping,3well}:
$\hat O \sum_{\vec n} C_{\vec n} |\vec n;t\rangle \equiv 
\sum_{\vec n} C^{\hat O}_{\vec n} |\vec n;t\rangle$.
The second summation means
that the vector of changed time-dependent coefficients 
$\C^{\hat O}(t)=\{C^{\hat O}_{\vec n}\}$
runs over {\it the same} time-dependent configurations
generated by distributing the $N$ identical particles over $M$ orbitals.
Below we treat the operation of one-body and two-body operators $\hat O$.
When needed and for the uniformity of the presentation, 
we exploit this notation also for the multiplication 
by a constant.
For simplicity, 
we denote in Eqs.~(\ref{MCTDHX_equ}) and (\ref{local}) 
the interaction $\hat W(\r-\r')$ such
that it depends, 
as is common, 
on the distance between the (identical) particles.
The more general case of interaction between identical particles 
which is merely symmetric in the coordinates of the particles is implicitly included.
Spin degrees-of-freedom are suppressed and implicit summation
on them 
is assumed throughout this work 
and to be unambiguously performed.

The differential condition is satisfied by Eq.~(\ref{MCTDHX_equ}) \cite{unified} 
and given by:
\beq\label{diff_cond}
i\langle\phi_k|\dot\phi_q\rangle = 0, \qquad k,q=1,\ldots,M.
\eeq
This guarantees the orthonormality of the orbitals at all times:
\beq\label{ortho_cond}
\langle\phi_k|\phi_q\rangle = \delta_{kq}, \qquad k,q=1,\ldots,M.
\eeq
The projector is given by:
\beq\label{projector}
\hat {\mathbf P} = 1 - \sum_{j'=1}^{M} \left|\phi_{j'}\right>\left<\phi_{j'}\right|.  
\eeq
Finally, the time evolution 
of the coefficients is, of course, unitary which guarantees
their normalization:
\beq\label{C_ortho_cond}
 \C^\dag\C = \1 .
\eeq 

Except for a unified and practical way of writing in terms of reduced
density matrices \cite{unified},
equations of motion (\ref{MCTDHX_equ}) 
provide the standard representation of MCTDHX theories.
They describe the evolution of the orbitals in their orthogonal space 
[in view of the differential condition 
Eq.~(\ref{diff_cond})] 
and the unitary evolution of the expansion coefficients.
It turns out that with a single and simple unitary transformation one can achieve 
orthogonal propagation 
for the coefficients' time evolution 
as well.
This result stands for itself and will also prove to be instrumental for linear response.
Explicitly, with the assignment of a joint 
time-dependent phase to all coefficients:
\beq\label{P_MCTDHX_phase}
 \C \longrightarrow \C e^{-i \int^t dt' \C^\dag(t') \H(t') \C(t')},
\eeq
the ``fully-projected" equations of motion of MCTDHX take on the novel form:
\beqn\label{P_MCTDHX_equ}
& &  \hat {\mathbf P} \sum_{q=1}^M [\rho_{kq} \hat h +
 \sum^M_{s,l=1}\rho_{kslq} \hat{W}_{sl}] |\phi_q\rangle = i \sum_{q=1}^M \rho_{kq} |\dot\phi_q\rangle, \qquad k=1,\ldots,M, \ \\
& & {\mathbf P}_C{\mathbf H}(t)\C(t) = i\frac{\partial \C(t)}{\partial t} 
\quad \Longleftrightarrow \quad {\mathbf P}_C\C^{\hat H}(t) = i \dot \C(t) \quad
\Longleftrightarrow \quad 
\C^{\hat H - \C^\dag \C^{\hat H}}(t) = i \dot \C(t), \nonumber \
\eeqn
with the coefficients' projector operator given by:
\beq\label{C_projector}
 {\mathbf P}_C = \1 - \C\C^\dag.
\eeq
Now, also the coefficients' part satisfies a differential condition:
\beq\label{C_diff_cond}
i\C^\dag\dot\C = 0.
\eeq
Consequently and appealingly,
by combining Eq.~(\ref{diff_cond}) for the orbitals and Eq.~(\ref{C_diff_cond}) for the expansion coefficients
it is seen that the MCTDHX wavefunction $|\Psi(t)\rangle = \sum_{\vec n} C_{\vec n} |\vec n;t\rangle$
evolves in a completely orthogonal manner,
namely $i\langle\Psi(t)|\dot\Psi(t)\rangle = 0$.
This is just as the time-dependent wavefunction of the ``projected'' 
Schr\"odinger equation does,
see Eq.~(\ref{SE_5}).

We will also be needing the static (time-independent) theories of MCTDHX --
MCHX (X=B,F), see Refs.~\cite{MCHB,MCHF1,MCHF2}.
This is because, as mentioned above, 
the linear response is to be performed around a static solution
of the many-particle system, 
typically the ground state.
The MCHX equations are given, 
e.g., by using imaginary-time propagation on MCTDHX, 
by:
\beqn\label{MCHX_equ}
& & \hat {\mathbf P} \sum_{q=1}^M [\rho_{kq} \hat h +
 \sum^M_{s,l=1}\rho_{kslq} \hat{W}_{sl}] |\phi_q\rangle = 0, \qquad k=1,\ldots,M, 
\qquad \Longleftrightarrow \  \\
& & \sum_{q=1}^M [\rho_{kq} \hat h - \mu_{kq} +
 \sum^M_{s,l=1}\rho_{kslq} \hat{W}_{sl}] |\phi_q\rangle = 0,  \qquad k=1,\ldots,M, \nonumber \\
 & &  \mu_{kq} = \sum^M_{j=1} \langle\phi_q|[\rho_{kj} \hat h + \sum^M_{s,l=1}\rho_{kslj} \hat{W}_{sl}]|\phi_j\rangle,  
\qquad k,q=1,\ldots,M, \nonumber \\
& & {\mathbf P}_C {\mathbf H} \C = \0 \ \Longleftrightarrow \ 
     {\mathbf H} \C = \varepsilon \C \Longleftrightarrow \
     \C^{\hat H - \C^\dag \C^{\hat H}} = 0 \Longleftrightarrow \  
     \C^{\hat H-\varepsilon} = \0 , \ \ \ \ 
     H_{\vec{n}\vec{n}'} = \langle\vec{n}|\hat H|\vec{n}'\rangle. \nonumber \
\eeqn
Note that in MCHX the matrix of Lagrange multipliers $\{\mu_{kq}\}$ 
is hermitian and can be diagonalized, namely
\beq\label{mu_diag}
  \mu_{kq}, \quad k,q=1,\ldots,M \qquad  \longrightarrow  
\qquad \mu_{k}, \qquad k=1,\ldots,M.
\eeq
We also recall that the eigenenergy $\varepsilon$ is, 
in fact, a Lagrange multiplier that ensures 
(redundantly -- in the time-dependent case) the vector 
of coefficients $\C$ to be normalized.

In the linear-response derivation that follows and whenever unambiguous, 
we denote the time-dependent 
and time-independent quantities by the same symbols,
to avoid cumbersome notation.

%%%%%%%%%%%%%%%%%%%%%%%%%%%%%%%%%%%%%%%%%%%%%%%%
\subsection{Perturbation and variation: Orthogonality}\label{Pert_var_orth_sec}
%%%%%%%%%%%%%%%%%%%%%%%%%%%%%%%%%%%%%%%%%%%%%%%%

We derive the linear-response 
theory from MCTDHX using a small perturbation {\it around} 
the MCHX solution, typically the ground state.
Thus, we have the following ansatz for the perturbing fields and the perturbed wavefunction:
\beqn\label{LR_ansatz}
& & \phi_k(\r,t) \approx \phi_k(\r) + \delta\phi_k(\r,t), \qquad k=1,\ldots,M, \nonumber \\
& & \delta\phi_k(\r,t) = u_k(\r) e^{-i \omega t} + v^\ast_k(\r) e^{+i \omega t},  \qquad k=1,\ldots,M, \nonumber \\
& & \C(t) \approx [\C + \delta\C(t)] \quad \Longleftrightarrow \quad 
\C(t) \approx e^{-i \varepsilon t} [\C + \delta\C(t)] 
\quad \mathrm{(without \ coefficients' \ projector)}, \nonumber \\
& & \delta\C(t) = \C_u e^{-i \omega t} + \C^\ast_v e^{+i \omega t}, \nonumber \\
& & \delta \hat h(\r,t) = \hat f^\dag(\r) e^{-i \omega t} + \hat f(\r) e^{+i \omega t}, \nonumber \\
& & \delta \hat W(\r-\r',t) = \hat g^\dag(\r-\r') e^{-i \omega t} + \hat g(\r-\r') e^{+i \omega t}. \
\eeqn
Here, $\{\delta\phi_k(\r,t)\}$ and $\delta\C(t) = \{\delta C_{\vec n}(t)\}$
are the perturbed parts of the orbitals and
coefficients, respectively, 
and comprised of $u$ and $v$ parts.
The operators $\hat f(\r)$ and $\hat g(\r-\r')$ generate one-body and two-body perturbations.
We consider a time-dependent perturbation hence $\omega>0$. 
The Hamiltonian including the perturbation is hermitian.

As mentioned above, we use in this work the same symbols for time-dependent quantities
and their zeroth-order time-independent parts when unambiguous.
It is to be understood that all relations containing perturbation parts
are linear with respect to the latter 
and generally hold to first order only
(as it should be per definition for 
a linear-response theory). 

Substituting the first expansion in Eq.~(\ref{LR_ansatz}) into the
orthonormality property 
of the orbitals, Eq.~(\ref{ortho_cond}), 
we obtain (to first order) the `integration-by-parts' relation:
\beq\label{integ_part_cond}
\langle\delta\phi_k(\r,t)|\phi_q(\r)\rangle = - \langle\phi_k(\r)|\delta\phi_q(\r,t)\rangle, \qquad k,q=1,\ldots,M.
\eeq
Similarly, making use of the differential condition Eq.~(\ref{diff_cond}) of MCTDHX one 
immediately arrives at the orthogonality of the perturbed parts of the orbitals
with respect to the ground-state manifold of orbitals, namely:
\beq\label{pert_orb_orth}
 \langle\phi_k(\r)|u_q(\r)\rangle = 0, \quad  \langle\phi_k(\r)|v^\ast_q(\r)\rangle = 0,  \qquad k,q=1,\ldots,M.
\eeq 
This relation is obtained to first order for $\omega > 0$
\footnote{Otherwise, i.e., for $\omega =0$,
the perturbation itself is time-independent, 
the ansatz for the variation
[see second line of Eq.~(\ref{LR_ansatz})]
is also time-independent, 
and hence Eq.~(\ref{diff_cond})
is satisfied automatically. 
This specific case which is ``in-resonance" with the 
zero-mode excitations 
{\it formally} requires a separate derivation in which the starting 
point is the response to a time-independent perturbation. 
However, since we are generally interested in excitations with 
a non-vanishing frequency (energy), 
we will not pursue this matter further here.}.
Any relation between the perturbed parts of the orbitals 
themselves is to second order,
and cannot be deduced from the differential condition 
Eq.~(\ref{diff_cond}) within linear response.

Analogous relations can be obtained
for the expansion coefficients.
Substituting the third expansion (either of the two) 
in Eq.~(\ref{LR_ansatz}) into the
orthonormality property of the coefficients, 
Eq.~(\ref{C_ortho_cond}), 
we obtain (to first order) the `integration-by-parts' relation:
\beq\label{C_integ_part_cond}
 (\delta\C)^\dag \C = - \C^\dag \delta\C.
\eeq
Similarly,
making use of the differential condition for the coefficients, 
Eq.~(\ref{C_diff_cond}), one 
immediately arrives at the orthogonality of the perturbed parts of the coefficients 
with respect to the ground-state (vector of) coefficients:
\beq\label{pert_C_orth}
 \C^\dag \C_u = 0, \quad  \C^\dag \C^\ast_v = 0.
\eeq 
Again, this relation is obtained to first order for $\omega > 0$
\footnote{Utilizing the 
differential-condition-derived
orthogonality relations
of the response amplitudes,
Eqs.~(\ref{pert_orb_orth}) and (\ref{pert_C_orth}), 
renders the `integration-by-parts' relations, Eqs.~(\ref{integ_part_cond}) and (\ref{C_integ_part_cond}), trivial.
Nonetheless, we find it instrumental to explicitly employ also the latter
in the derivation of the linear-response equations in Subsec.~\ref{Orbitals_sec},
also see Ref.~\cite{LR_MCTDHB}.}.

%%%%%%%%%%%%%%%%%%%%%%%%%%%%%%%%%%%%%%%%%%%%%%%%
\subsection{0-th order: MCHB and MCHF}\label{0th_order_sec}
%%%%%%%%%%%%%%%%%%%%%%%%%%%%%%%%%%%%%%%%%%%%%%%%

The linear-response 
equations are derived by substituting 
the ansatz Eq.~(\ref{LR_ansatz}) into the MCTDHX equations of motion.
To zeroth order one simply obtains 
the MCHX system itself which we repeat for reference:
\beqn\label{LR_0th}
& & \hat {\mathbf P} \sum_{q=1}^M [\rho_{kq} \hat h +
 \sum^M_{s,l=1}\rho_{kslq} \hat{W}_{sl}] |\phi_q\rangle = 0, \qquad k=1,\ldots,M, \nonumber \\
& & {\mathbf P}_C{\mathbf H}\C = \0 \qquad \Longleftrightarrow
\qquad \C^{\hat H-\varepsilon} =  \0 . \
\eeqn
When possible or instructive, 
we will avoid writing the Lagrange multipliers explicitly.

%%%%%%%%%%%%%%%%%%%%%%%%%%%%%%%%%%%%%%%%%%%%%%%%
\subsection{1-st order: Orbitals}\label{Orbitals_sec}
%%%%%%%%%%%%%%%%%%%%%%%%%%%%%%%%%%%%%%%%%%%%%%%%

We now derive separately the first order (linear-response) equations emanating
from the orbitals' and from the coefficients' (subsequent Subsec.~\ref{Coefficients_sec}) 
parts of MCTDHX.
This is done by formally taking the variation of
the MCTDHX equations of motion.
We find from Eq.~(\ref{MCTDHX_equ}) 
for the perturbed orbitals that:
\beqn\label{LR_1st_orb_1}
& &  \delta\hat {\mathbf P} \sum_{q=1}^M [\rho_{kq} \hat h +
 \sum^M_{s,l=1}\rho_{kslq} \hat{W}_{sl}] |\phi_q\rangle + \nonumber \\
& & + \hat {\mathbf P} \sum_{q=1}^M [\delta \rho_{kq} \hat h + \rho_{kq} \delta \hat h +
 \sum^M_{s,l=1} (\delta \rho_{kslq} \hat{W}_{sl} + \rho_{kslq} \delta\hat{W}_{sl})] |\phi_q\rangle + \nonumber \\
& & + \hat {\mathbf P} \sum_{q=1}^M [\rho_{kq} \hat h +
 \sum^M_{s,l=1}\rho_{kslq} \hat{W}_{sl}] | \delta\phi_q\rangle = \nonumber \\
& & = i \sum_{q=1}^M \rho_{kq} |\delta\dot\phi_q\rangle, \qquad k=1,\ldots,M.  \
\eeqn
Note that a term with $\delta \rho_{kq}$ 
in front of the time derivative is to second order
and hence does not enter here.

The evaluation of the first term in Eq.~(\ref{LR_1st_orb_1}) 
employs the derived `integration-by-parts' 
relation Eq.~(\ref{integ_part_cond}).
Thus we find:
\beq\label{LR_1st_orb_2}
 \delta\hat {\mathbf P} \sum_{q=1}^M [\rho_{kq} \hat h +
 \sum^M_{s,l=1}\rho_{kslq} \hat{W}_{sl}] |\phi_q\rangle 
= - \hat {\mathbf P} \sum_{q=1}^M \mu_{kq} |\delta \phi_q\rangle, \qquad k=1,\ldots,M. 
\eeq
Using Eq.~(\ref{pert_orb_orth}) and similarly to Eq.~(\ref{LR_1st_orb_2}),
the projection operator $\hat {\mathbf P}$ can also be reinstated in 
front of the
time derivative on the right-hand side 
of the linear-response system Eq.~(\ref{LR_1st_orb_1}).
This will be used later on.
With Eq.~(\ref{LR_1st_orb_2}), 
the variation in Eq.~(\ref{LR_1st_orb_1}) simplifies and we get:
\beqn\label{LR_1st_orb_3}
& &  \hat {\mathbf P} \sum_{q=1}^M [\delta \rho_{kq} \hat h + \rho_{kq} \delta \hat h +
 \sum^M_{s,l=1} (\delta \rho_{kslq} \hat{W}_{sl} + \rho_{kslq} \delta\hat{W}_{sl})] |\phi_q\rangle + \nonumber \\
& & + \hat {\mathbf P} \sum_{q=1}^M [\rho_{kq} \hat h - \mu_{kq} +
 \sum^M_{s,l=1}\rho_{kslq} \hat{W}_{sl}] | \delta\phi_q\rangle = \nonumber \\
& & = i \sum_{q=1}^M \rho_{kq} |\delta\dot\phi_q\rangle, \qquad k=1,\ldots,M.  \
\eeqn
Eq.~(\ref{LR_1st_orb_3}) is the generic form for the linear-response 
equations of the orbitals, 
{\it formally} for $\omega > 0$. 
The derivation for time-independent perturbations is beyond our scope here.
In case of bosons and for a single orbital, 
that is for $M=1$, 
Eq.~(\ref{LR_1st_orb_3}) boils down to
a particle-conserving version of 
the Bogoliubov--de Gennes equations \cite{Castin_Dum}.
When the number of fermions equals the number of orbitals,
namely $M=N$,
Eq.~(\ref{LR_1st_orb_3}) boils down to
the random-phase approximation, see in this context, e.g., Ref.~\cite{Ripka_book}.
The orbitals' projector guarantees that this is an orthogonal version of the random-phase approximation, 
where all orbitals' perturbations are automatically orthogonal to the ground-state determinant.

Let us now proceed and derive the explicit equations for the perturbed orbitals,
namely substitute the 
ansatz Eq.~(\ref{LR_ansatz}) 
into the result Eq.~(\ref{LR_1st_orb_3}).
The first-order equation associated with $e^{-i \omega t}$ 
is given by:
\beqn\label{LR_1st_orb_4}
& &  \hat {\mathbf P} \sum_{q=1}^M [\delta \rho_{kq}|_{e^{- i\omega t}} \hat h + 
 \sum^M_{s,l=1} (\delta \rho_{kslq}|_{e^{- i\omega t}} \hat{W}_{sl} + \rho_{kslq} \delta\hat{W}_{sl}|_{e^{- i\omega t}})] |\phi_q\rangle + \nonumber \\
& & + \hat {\mathbf P} \sum_{q=1}^M [\rho_{kq} (\hat h -\omega) - \mu_{kq} +
 \sum^M_{s,l=1}\rho_{kslq} \hat{W}_{sl}] |u_q\rangle  
= \nonumber \\
& & = - \hat {\mathbf P} \sum_{q=1}^M 
(\rho_{kq} \hat f^\dag + \sum^M_{s,l=1} \rho_{kslq} \{\hat g^\dag\}_{sl})|\phi_q\rangle, \qquad k=1,\ldots,M,  \
\eeqn
where $\{\hat g^\dag\}_{sl} = 
\int d\r' \phi^\ast_s(\r') \hat g^\dag(\r-\r') \phi_l(\r')$.
Since in view of the above it is allowed, 
we for now put the $\omega$-term inside, i.e., under the projector.

We can now proceed and substitute 
the perturbed quantities under the variation in Eq.~(\ref{LR_1st_orb_4}).
This leads to 
the following ingredients for 
the $e^{-i \omega t}$ equation:
\beqn\label{del_e_minus_1}
  \hat{W}_{sl} & & \qquad \Longrightarrow \qquad  \delta\hat{W}_{sl}|_{e^{- i\omega t}} = \hat{W}_{l^\ast v_s} + \hat{W}_{su_l}, \nonumber \\
  \rho_{kq} = \C^\dag \cdot \C^{\hat \rho_{kq}}  & & \qquad \Longrightarrow \qquad  
\delta\rho_{kq}|_{e^{- i\omega t}} = (\C^{\hat \rho_{kq}})^t \cdot \C_v + (\C^{\hat \rho_{qk}})^\dag \cdot \C_u, \nonumber \\
  \rho_{kslq} = \C^\dag \cdot \C^{\hat \rho_{kslq}} & & \qquad \Longrightarrow \qquad 
\delta\rho_{kslq}|_{e^{- i\omega t}} = (\C^{\hat \rho_{kslq}})^t \cdot \C_v + (\C^{\hat \rho_{qlsk}})^\dag \cdot \C_u, \
\eeqn
where, to remind,
the transformed coefficients $\C^{\hat \rho_{kq}}=\{C_{\vec n}^{\hat \rho_{kq}}\}$ are defined as
$\hat \rho_{kq} \sum_{\vec n} C_{\vec n} |\vec n;t\rangle \equiv 
\sum_{\vec n} C_{\vec n}^{\hat \rho_{kq}} |\vec n;t\rangle$,
and similarly for $\C^{\hat \rho_{kslq}} = \{C_{\vec n}^{\hat \rho_{kslq}}\}$.
The final result reads:
\beqn\label{LR_1st_orb_6}
& &  \hat {\mathbf P} \sum_{q=1}^M [\{(\C^{\hat \rho_{kq}})^t \cdot \C_v 
+ (\C^{\hat \rho_{qk}})^\dag \cdot \C_u\} \hat h + 
 \sum^M_{s,l=1} \{(\C^{\hat \rho_{kslq}})^t \cdot \C_v 
+ (\C^{\hat \rho_{qlsk}})^\dag \cdot \C_u \} \hat{W}_{sl}] |\phi_q\rangle + \nonumber \\
& & + \hat {\mathbf P} \sum_{q=1}^M [\rho_{kq} (\hat h -\omega) - \mu_{kq} +
 \sum^M_{s,l=1}\rho_{kslq} (\hat{W}_{sl} \pm \hat{K}_{sl})]|u_q\rangle + \hat {\mathbf P} \sum_{q,s,l=1}^M
\rho_{kqls} \hat{K}_{l^\ast s} |v_q\rangle 
= \nonumber \\
& & = - \hat {\mathbf P} \sum_{q=1}^M 
(\rho_{kq} \hat f^\dag + \sum^M_{s,l=1} \rho_{kslq} \{\hat g^\dag\}_{sl})|\phi_q\rangle, \qquad k=1,\ldots,M,  \
\eeqn
where the $\pm$ sign refers to bosons or fermions, respectively, 
and the exchange operator is introduced and defined as: 
\beq\label{exchange}
 \hat K_{sl} = \int d\r' \phi^\ast_s(\r') \hat W(\r-\r') \hat {\mathcal P}_{\r\r'} \phi_l(\r'), \qquad 
 \hat K_{sl} f(\r) \equiv  \hat W_{sf} \phi_l(\r)
\eeq
with $\hat {\mathcal P}_{\r\r'}$ permuting 
the coordinates of two particles. 
We emphasize that MCTDHX and MCHX are formulated with {\it local} potentials only, see Eq.~(\ref{local}).
For their linear-response theories, 
the exchange potentials, Eq.~(\ref{exchange}), 
with the perturbed orbitals appear,
see the term $\pm \hat{K}_{sl}|u_q\rangle$ in Eq.~(\ref{LR_1st_orb_6}).
Furthermore, the $\pm$ in front of this exchange term
implies that the dependence on the
particles' statistics appears explicitly in linear response,
also see further below,
unlike the implicit dependence (``invariance") on particles' statistics of 
MCTDHX and MCHX, see Ref.~\cite{unified}.

The first-order equation associated with $e^{+i \omega t}$ is given 
by:
\beqn\label{LR_1st_orb_5}
& &  \hat {\mathbf P}^\ast \sum_{q=1}^M [\{\delta \rho_{kq}|_{e^{+ i\omega t}}\}^\ast \hat h^\ast + 
 \sum^M_{s,l=1} (\{\delta \rho_{kslq}|_{e^{+ i\omega t}}\}^\ast \hat{W}_{ls}
 + \rho_{qlsk} \{\delta\hat{W}_{sl}|_{e^{+ i\omega t}}\}^\ast)] |\phi^\ast_q\rangle + \nonumber \\
& & + \hat {\mathbf P}^\ast \sum_{q=1}^M [\rho_{qk} (\hat h^\ast +\omega) - \mu_{qk} +
 \sum^M_{s,l=1}\rho_{qlsk} \hat{W}_{ls}] |v_q\rangle
= \nonumber \\
& & = - \hat {\mathbf P}^\ast \sum_{q=1}^M 
(\rho_{qk} \hat f^\ast + \sum^M_{s,l=1} \rho_{qlsk} \{\hat g^\dag\}_{ls})|\phi^\ast_q\rangle, \qquad k=1,\ldots,M,  \
\eeqn
where the hermiticity of the Lagrange 
multipliers' matrix $\mu^\ast_{kq}=\mu_{qk}$ has been used.
Again, we for now put the $\omega$-term under the projector
in view of Eq.~(\ref{pert_orb_orth}).

Similarly to the above treatment 
we obtain the following ingredients 
for the $e^{+i\omega t}$ equation 
[utilizing the $e^{-i\omega t}$ ones, see Eq.~(\ref{del_e_minus_1})]:
\beqn\label{del_e_plus_1}
  {[\delta \hat{W}_{sl}|_{e^{+ i\omega t}}]}^\dag & & \qquad \Longleftrightarrow \qquad  \delta\hat{W}_{ls}|_{e^{- i\omega t}}, \nonumber \\
  {[\delta \rho_{kq} |_{e^{+ i\omega t}}]}^\dag & &  \qquad \Longleftrightarrow \qquad \delta\rho_{qk}|_{e^{- i\omega t}}, \nonumber \\
  {[\delta\rho_{kslq}|_{e^{+ i\omega t}}]}^\dag & & \qquad \Longleftrightarrow \qquad  \delta\rho_{qlsk}|_{e^{- i\omega t}}, \
\eeqn
where, of course, 
${[\delta \hat h|_{e^{+ i\omega t}}]}^\dag = \delta\hat h|_{e^{- i\omega t}}$ holds
\footnote{In case the interparticle 
interactions contain momenta operators,
care should be exercised in replacing $\ast$ by $\dag$ atop 
an operator like $\delta\hat{W}_{ls}$.
Non-local perturbing fields are beyond
the scope of the present work.}.
The final result reads:
\beqn\label{LR_1st_orb_7}
& &  \hat {\mathbf P}^\ast \sum_{q=1}^M
[\{(\C^{\hat \rho_{qk}})^t \cdot \C_v + (\C^{\hat \rho_{kq}})^\dag \cdot \C_u\}
 \hat h^\ast + 
 \sum^M_{s,l=1}
\{(\C^{\hat \rho_{qlsk}})^t \cdot \C_v + (\C^{\hat \rho_{kslq}})^\dag \cdot \C_u\}
\hat{W}_{ls}] |\phi^\ast_q\rangle + \nonumber \\
& & + \hat {\mathbf P}^\ast \sum_{q=1}^M [\rho_{qk} (\hat h^\ast +\omega) - \mu_{qk} +
 \sum^M_{s,l=1}\rho_{qlsk} (\hat{W}_{ls} 
\pm \hat{K}_{s^\ast l^\ast})] |v_q\rangle 
+ \hat {\mathbf P}^\ast \sum_{q,s,l=1}^M
 \rho_{slqk} \hat{K}_{ls^\ast} |u_q\rangle
= \nonumber \\
& & = - \hat {\mathbf P}^\ast \sum_{q=1}^M 
(\rho_{qk} \hat f^\ast + \sum^M_{s,l=1} \rho_{qlsk} 
\{\hat g^\dag\}_{ls})|\phi^\ast_q\rangle, \qquad k=1,\ldots,M.  \
\eeqn
Observe and compare Eqs.~(\ref{LR_1st_orb_7}) and (\ref{LR_1st_orb_6}) 
for how they are related.
We will return to this matter in Subsec.~\ref{matrix_form_sec}.

%%%%%%%%%%%%%%%%%%%%%%%%%%%%%%%%%%%%%%%%%%%%%%%%
\subsection{1-st order: Coefficients}\label{Coefficients_sec}
%%%%%%%%%%%%%%%%%%%%%%%%%%%%%%%%%%%%%%%%%%%%%%%%

Next, we now move to the perturbed coefficients.
We find from Eq.~(\ref{MCTDHX_equ}) for the perturbed coefficients that:
\beq\label{LR_1st_coeff_1}
\delta\C^{\hat H} + \C^{\delta\hat H} = \varepsilon \delta\C + i\delta\dot\C \qquad \Longleftrightarrow \qquad
  \delta\C^{\hat H - \varepsilon} + \C^{\delta\hat H} = i\delta\dot\C.
\eeq
Furthermore, 
from the ``fully projected" 
MCTDHX equations of motion, 
Eq.~(\ref{P_MCTDHX_equ}), 
we find for
the perturbed coefficients that:
\beq\label{P_LR_1st_coeff_1}
\delta{\mathbf P}_C \C^{\hat H} + {\mathbf P}_C (\delta\C^{\hat H} + \C^{\delta\hat H}) =  i\delta\dot\C \qquad \Longleftrightarrow \qquad
  {\mathbf P}_C (\delta\C^{\hat H - \varepsilon} + \C^{\delta\hat H}) = i\delta\dot\C,
\eeq 
where use has been made in the relation [compare to Eq.~(\ref{LR_1st_orb_2})]:
\beq\label{C_pro_inter_by_part}
\delta{\mathbf P}_C \H \C = -\varepsilon {\mathbf P}_C \delta\C
\qquad \Longleftrightarrow \qquad
\delta{\mathbf P}_C \C^{\hat H} =  {\mathbf P}_C \delta\C^{- \varepsilon}.
\eeq
From now on we will use 
the ``fully projected'' MCTDHX theory
in order to derive the coefficients' equation of motion.
The results from the standard
MCTDHX equations of motion
can generally be obtained from the following 
by dropping ${\mathbf P}_C$ therein.

We will also be needing the Hamiltonian and its variation, 
expressed in terms of density operators
\footnote{Strictly speaking, 
within the Fock-space mapping for
identical-particle configurations, Ref.~\cite{Mapping}, 
the upper summation $M$ is required only for the
creation operators, but it is convenient
to leave it also for the annihilation operators.}:
\beq\label{Ham_den}
 \hat H = \sum_{k,q=1}^{M} 
[h_{kq} \hat\rho_{kq} + \frac{1}{2} \sum_{s,l=1}^{M} W_{ksql} \hat\rho_{kslq}]
\eeq
and
\beqn\label{del_Ham_den}
& & \delta \hat H = \sum_{k,q=1}^{M} [(h_{\delta kq} + h_{k\delta q} 
+ \{\delta h\}_{kq}) \hat\rho_{kq} + \\
& & + \frac{1}{2} \sum_{s,l=1}^{M} 
(W_{\delta ksql} + W_{k\delta sql} + W_{ks\delta ql} + W_{ksq\delta l}
+ \{\delta W\}_{ksql}) \hat\rho_{kslq}] = \nonumber \\
& & = \sum_{k,q=1}^{M} 
[(\{h^\ast\}_{q^\ast \delta k^\ast} + h_{k\delta q} + \{\delta h\}_{kq}) \hat\rho_{kq} 
+ \sum_{s,l=1}^{M} 
(W_{sq^\ast l\delta k^\ast} + W_{skl\delta q} + 
\frac{1}{2} \{\delta W\}_{ksql}) \hat\rho_{kslq}]. \nonumber \
\eeqn
Note that for current convenience  
we write the perturbed part of the Hamiltonian
(i.e., the perturbing fields) 
only in its variation part, 
also see below.

Let us now proceed and derive the explicit equations for the perturbed coefficients,
namely substitute the ansatz Eq.~(\ref{LR_ansatz}) into the result Eq.~(\ref{P_LR_1st_coeff_1}).
The first-order equation associated with $e^{-i \omega t}$ is given by:
\beqn\label{LR_1st_coeff_2}
& & {\mathbf P}_C\C_u^{\omega + \varepsilon - \hat H} = 
{\mathbf P}_C\C^{\delta\hat H|_{e^{- i\omega t}}}, \\
& &   \delta\hat H|_{e^{- i\omega t}} = 
\sum_{k,q=1}^{M} [( \{h^\ast\}_{q^\ast v_k} + h_{ku_q} + \{f^\dag\}_{kq}) \hat\rho_{kq} 
+ \sum_{s,l=1}^{M} (W_{sq^\ast lv_k} + W_{sklu_q} + 
\frac{1}{2}\{g^\dag\}_{ksql}) \hat\rho_{kslq}]. \nonumber \
\eeqn
The first-order equation associated with $e^{+i \omega t}$ is given by:
\beqn\label{LR_1st_coeff_3}
& & \!\!\! {\mathbf P}^\ast_C \C_v^{\varepsilon - \omega - \hat H^\star} = 
{\mathbf P}^\ast_C (\C^\ast)^{\{\delta\hat H|_{e^{+ i\omega t}}\}^\star}, \\
& & \!\!\! \hat H^\star = \sum_{k,q=1}^{M} 
[h_{qk} \hat\rho_{kq} + \frac{1}{2} \sum_{s,l=1}^{M} 
W_{lqsk} \hat\rho_{kslq}], \nonumber \\
& & \!\!\!  \{\delta\hat H|_{e^{+ i\omega t}}\}^\star = 
\sum_{k,q=1}^{M} [(h_{qu_k} + \{h^\ast\}_{k^\ast v_q} + \{f^\dag\}_{qk}) \hat\rho_{kq} 
 + \sum_{s,l=1}^{M} (W_{lqsu_k} + W_{lk^\ast sv_q} 
+ \frac{1}{2}\{g^\dag\}_{lqsk}) \hat\rho_{kslq}]. \nonumber \
\eeqn
Here $\varepsilon$ is the MCHX ground-state energy and is, of course, real valued.
Note that the complex conjugate ${\mathbf P}^\ast_C$ appears for the perturbed coefficients 
as does the complex conjugate $\hat {\mathbf P}^\ast$ for the perturbed orbitals,
see Eq.~(\ref{LR_1st_orb_7}).
In Eq.~(\ref{LR_1st_coeff_3}) we introduced for convenience the star ($\star$) 
of a second-quantized operator, $\hat O^\star$,
which is related 
to the original operator $\hat O$ as follows:
(i) take the complex conjugate of the one-body and two-body matrix elements with respect to the orbitals, and
(ii) do not take the hermitian conjugate of the density operators.
The star operation comes naturally when utilizing complex conjugation within 
mapping in Fock space.
Of course, since the Hamiltonian is hermitian, 
the usual hermiticity 
holds for the $e^{\mp i \omega t}$ variations of the Hamiltonian:
\beq\label{hermit_Coeff_pm}
 \{\delta\hat H|_{e^{+ i\omega t}}\}^\dag =  \{\delta\hat H|_{e^{- i\omega t}}\}.
\eeq
We remind the notation for the 
transformed coefficients in Eqs.~(\ref{LR_1st_coeff_2}) and (\ref{LR_1st_coeff_3})
where, e.g.,
$\C_v^{\varepsilon - \omega - \hat H^\star} = \{C_{v, \vec n}^{\varepsilon - \omega - \hat H^\star}\}$ 
is defined as follows:
$ (\varepsilon - \omega - \hat H^\star)  \sum_{\vec n} C_{v, \vec n} |\vec n;t\rangle \equiv 
\sum_{\vec n} C_{v, \vec n}^{\varepsilon - \omega - \hat H^\star} |\vec n;t\rangle$,
with $\C_v = \{C_{v, \vec n}\}$.

%%%%%%%%%%%%%%%%%%%%%%%%%%%%%%%%%%%%%%%%%%%%%%%%%
\subsection{The linear-response matrix system and its formal solution}\label{matrix_form_sec}
%%%%%%%%%%%%%%%%%%%%%%%%%%%%%%%%%%%%%%%%%%%%%%%%%

We now proceed and cast the coupled linear-response system,
Eqs.~(\ref{LR_1st_orb_6}), (\ref{LR_1st_orb_7}),
(\ref{LR_1st_coeff_2}), and (\ref{LR_1st_coeff_3}),
into a matrix form.
This is done in Subsec.~\ref{matrix_form_sec_s1}
where we first post the main result 
and subsequently 
go through the individual
ingredients that make it.
With the linear-response matrix system
we can formally solve the time-dependent many-body Schr\"odinger equation in linear
response.
This is done in Subsec.~\ref{matrix_form_sec_s2} 
where we first discuss the 
symmetry and other
properties of the
linear-response matrix system, 
and subsequently
solve for the perturbed time-dependent orbitals and coefficients,
and the MCTDHX wavefunction in linear response.

%%%%%%%%%%%%%%%%%%%%%%%%%%%%%%%%%%%%%%%%%%%%%%%%%
\subsubsection{Casting the linear-response equations into a matrix form}\label{matrix_form_sec_s1}
%%%%%%%%%%%%%%%%%%%%%%%%%%%%%%%%%%%%%%%%%%%%%%%%%

Since the responses 
of the orbitals and coefficients are coupled, 
the central framework is to look at the
system's response space as a combined
orbital--coefficient response space;
also see in this respect Ref.~\cite{LR_MCTDHB}.
Correspondingly, we define the combined vector of length
$2(M+N_{\mathit{conf}})$:
\beq\label{LR_vector}
\begin{pmatrix} 
 \u \\
 \v \\
 \C_u \\
 \C_v \\
\end{pmatrix},
\qquad \qquad 
\u = \{|u_q\rangle\}, \
\v = \{|v_q\rangle\}, \
q=1,\ldots,M
\eeq
which collects all response amplitudes 
together 
for a given perturbed wavefunction.

Within this combined orbital--coefficient response space,
the {\it final} 
result for the linear-response working matrix equation is:
\beq\label{LR_matrix_EO_working}
\left(\bcalL-\omega\right)
\begin{pmatrix} 
 \u \\
 \v \\
 \C_u \\
 \C_v \\
\end{pmatrix} =
\bcalR.
\eeq
The linear-response matrix $\bcalL$ is explicitly assembled below,
as does the vector $\bcalR$ which collects the perturbing fields,
see Eq.~(\ref{LR_ansatz}).
We call Eq.~(\ref{LR_matrix_EO_working}) LR-MCTDHX theory,
see in this context LR-MCTDHB \cite{LR_MCTDHB}.

To solve the LR-MCTDHX linear-response system Eq.~(\ref{LR_matrix_EO_working}),
and hence the Schr\"odinger equation in linear response,
we have to diagonalize the linear-response matrix $\bcalL$
and find its 
excitations energies $\{\omega_k\}$ and eigenvectors: 
\beq\label{LR_matrix_diag_1}
\bcalL
\begin{pmatrix} 
 \u^k \\
 \v^k \\
 \C_u^k \\
 \C_v^k \\
\end{pmatrix} =
 w_k
\begin{pmatrix} 
 \u^k \\
 \v^k \\
 \C_u^k \\
 \C_v^k \\
\end{pmatrix}
\equiv w_k \R^k
\eeq
This will be done in details in the following subsection \ref{matrix_form_sec_s2}
\footnote{The shorthand notation for the eigenvectors on the right-hand side of Eq.~(\ref{LR_matrix_diag_1}) is 
also chosen to point out 
that $\{\R^k\}$ are the right eigenvectors of $\bcalL$,
and should not be confused 
with the response vector 
in the linear-response Eq.~(\ref{LR_matrix_EO_working})
denoted by $\bcalR$.}.

We will now assemble and put together explicitly 
the associated response matrix.
The linear-response matrix is divided into four blocks:
\beq\label{LR_matrix}
 \bcalL =
\begin{pmatrix} 
 \bcalL_{oo} & \bcalL_{oc} \\
 \bcalL_{co} & \bcalL_{cc} \\
\end{pmatrix}
\eeq
and, like the response vector Eq.~(\ref{LR_vector}), 
its dimension is $2(M+N_{\mathit{conf}})$.
The structure of $\bcalL$ 
represents the fact that the linear-response
subspace is a combined space of orbitals and coefficients;
the size of $\bcalL$ is twice as large as 
their combined sizes.

Each of the four 
blocks of $\bcalL$
is divided itself into four 
sub-matrices, 
representing the $u$ and $v$ quantities.
Within each block, 
the four sub-matrices 
are linked between them
as will be seen below.
The orbital--orbital ($oo$) block reads:
\beq\label{LR_matrix_oo}
 \bcalL_{oo} =
\begin{pmatrix} 
 \bcalL_{oo}^u & \bcalL_{oo}^v \\
 -{(\bcalL_{oo}^v)}^\ast & -{(\bcalL_{oo}^u)}^\ast \\
\end{pmatrix},
\eeq
where ($k,q=1,\ldots,M$)
\beqn\label{LR_matrix_oo_1}
& & \!\!\!\! \bcalL_{oo}^u = \brho \hat h - \bmu + \bOmega \pm \bkappa^1, \nonumber \\
& &  \brho = \{\rho_{kq}\}, \ 
    \bmu = \{\mu_{kq}\}, \
    \bOmega = \{\Omega_{kq}\} = \left\{\sum^M_{s,l=1}\rho_{kslq} \hat{W}_{sl}\right\}, \
    \bkappa^1 = \{\kappa^1_{kq}\} = 
\left\{\sum^M_{s,l=1}\rho_{kslq} \hat{K}_{sl}\right\}, \nonumber \\
& & \!\!\!\! \bcalL_{oo}^v = \bkappa^2 = \{\kappa^2_{kq}\} = \left\{\sum_{s,l=1}^M
\rho_{kqls}\hat{K}_{l^\ast s}\right\}. \
\eeqn
Inspecting of the block $\bcalL_{oo}$ reveals that
the following further relations between its sub-matrices hold:
\beq\label{LR_matrix_oo_relation}
 (\bcalL_{oo}^u)^\dag = (\bcalL_{oo}^u), \qquad \qquad
 (\bcalL_{oo}^v)^{t} = (\bcalL_{oo}^v).
\eeq
The proof, especially with the non-local parts $\bkappa^1$ and $\bkappa^2$,
follows straightforwardly by multiplying
each sub-matrix with the identity 
(in the super-vector subspace of orbitals)
from its left and right sides.

The orbital--coefficient ($oc$) block reads:
\beq\label{LR_matrix_oc}
 \bcalL_{oc} =
\begin{pmatrix} 
 \bcalL_{oc}^u & \bcalL_{oc}^v \\
 -{(\bcalL_{oc}^v)}^\ast & -{(\bcalL_{oc}^u)}^\ast \\
\end{pmatrix},
\eeq
where ($k=1,\ldots,M$)
\beqn\label{LR_matrix_oc_1}
 & & \bcalL_{oc}^u =  
\left\{\sum_{q=1}^M [\hat h (\C^{\hat \rho_{qk}})^\dag +
  \sum^M_{s,l=1}\hat{W}_{sl} (\C^{\hat \rho_{qlsk}})^\dag]|\phi_q\rangle
\right\},
\nonumber \\
 & & \bcalL_{oc}^v =  
\left\{\sum_{q=1}^M [\hat h (\C^{\hat \rho_{kq}})^t +
 \sum^M_{s,l=1}\hat{W}_{sl} (\C^{\hat \rho_{kslq}})^t] 
|\phi_q\rangle
\right\}. \
\eeqn
The coefficient--orbital ($co$) block reads:
\beq\label{LR_matrix_co}
 \bcalL_{co} =
\begin{pmatrix} 
 \bcalL_{co}^u & \bcalL_{co}^v \\
 -{(\bcalL_{co}^v)}^\ast & -{(\bcalL_{co}^u)}^\ast \\
\end{pmatrix},
\eeq
where ($k=1,\ldots,M$)
\beqn\label{LR_matrix_co_1}
 & & \bcalL_{co}^u = \left\{\sum_{q=1}^M \langle \phi_q| 
[(\C^{\hat \rho_{qk}})\hat h +
 \sum^M_{s,l=1} (\C^{\hat \rho_{qlsk}}) \hat W_{ls}]
\right\}, \nonumber \\
 & & \bcalL_{co}^v = \left\{\sum_{q=1}^M \langle \phi_q^\ast| 
[(\C^{\hat \rho_{kq}})\hat h^\ast +
 \sum^M_{s,l=1} (\C^{\hat \rho_{kslq}}) \hat W_{sl}]
\right\}. \
\eeqn
Inspection the sub-matrices of
$\bcalL_{oc}$ and $\bcalL_{co}$
reveals
the following relation 
between the two off-diagonal
rectangular blocks of the response matrix $\bcalL$:
\beq\label{LR_matrix_oc_co_relation}
 (\bcalL_{oc}^u)^\dag = (\bcalL_{co}^u), \qquad \qquad
 (\bcalL_{oc}^v)^{t} = (\bcalL_{co}^v).
\eeq
The proof is visualized directly by multiplying
each sub-matrix with the identity 
(in the super-vector subspace of orbitals)
and the identity in the subspace of coefficients,
either from its left or right side,
respectively.
The relation in Eq.~(\ref{LR_matrix_oc_co_relation}) 
can be used to reduce the computational effort
in computing the response matrix,
since the two off-diagonal blocks of $\bcalL$
do not need to be computed independently
(see in this respect Ref.~\cite{LR_MCTDHB}).
Finally, the coefficient--coefficient 
($cc$) block reads:
\beq\label{LR_matrix_cc_mapping}
 \bcalL_{cc} =
\begin{pmatrix} 
 (\cdot)^{\hat H - \varepsilon} & {\0 }_c \\
  {\0 }_c & (\cdot)^{\varepsilon - \hat H^\star} \\
\end{pmatrix},
\eeq
where ${\0 }_c$ and ${\1 }_c$ (see below) are zero and unit 
matrices of dimension $N_{\mathit{conf}}$.
The $(\cdot)$ symbol means that the operation on the
response-coefficients' part is performed within the mapping
of coefficients \cite{Mapping},
and the $\star$ of the second-quantized Hamiltonian
is defined in Eq.~(\ref{LR_1st_coeff_3}).

We will also be needing the orbitals--coefficients
combined projector $\bcalP$ and combined metric $\bcalM$ matrices:
\beq\label{LR_matrix_P}
 \bcalP =
\begin{pmatrix} 
 \bcalP_o & {\0 }_{oc} \\
  {\0 }_{co} &  \bcalP_c \\
\end{pmatrix}, \qquad \qquad
 \bcalP_o =
\begin{pmatrix} 
 {\mathbf P} & {\0 }_o \\
  {\0 }_o &  {\mathbf P}^\ast \\
\end{pmatrix}, \
{\mathbf P} = \hat{\mathbf P} {\1 }_o,
\ \quad
 \bcalP_c =
\begin{pmatrix} 
 {\mathbf P}_C & {\0 }_c \\
  {\0 }_c & {\mathbf P}^\ast_C \\
\end{pmatrix},
\eeq
where ${\1 }_o$ and ${\0 }_o$ are unit and zero matrices of dimension $M$, 
${\0 }_{oc}$ is a zero rectangular matrix 
of dimension $M \times N_{\mathit{conf}}$, 
and ${\0 }_{co} = ({\0 }_{oc})^t$.
Similarly:
\beq\label{LR_matrix_M}
 \bcalM =
\begin{pmatrix} 
 \brho_o & {\0 }_{oc} \\
  {\0 }_{co} &  {\1 }_{2c} \\
\end{pmatrix}, \qquad \qquad
 \brho_o =
\begin{pmatrix} 
 \brho  & {\0 }_o \\
  {\0 }_o &   \brho^\ast \\
\end{pmatrix},
\eeq
where ${\1 }_{2c}$ is a unit matrix of dimension $2N_{\mathit{conf}}$.
Clearly, in view of 
the orthogonality relations Eqs.~(\ref{pert_orb_orth},\ref{pert_C_orth})
every response vector satisfies:
\beq\label{LR_P_vector}
\bcalP 
\begin{pmatrix} 
 \u \\
 \v \\
 \C_u \\
 \C_v \\
\end{pmatrix} =
\begin{pmatrix} 
 \u \\
 \v \\
 \C_u \\
 \C_v \\
\end{pmatrix},
\eeq
i.e., it lives in the complementary space of the 
ground-state wavefunction.

Finally, we collect the perturbing fields in the vector:
\beq\label{LR_perturbing_fields}
 \bcalM^{+\frac{1}{2}} \bcalR = \bcalM^{+\frac{1}{2}} 
\begin{pmatrix} 
 - \hat f^\dag \bphi \\
   \hat f^\ast \bphi^\ast \\
 - \C^{\left\{\sum_{k,q=1}^M \{f^\dag\}_{kq}\hat\rho_{kq}\right\}} \\
   (\C^\ast)^{\left\{\sum_{k,q=1}^M \{f^\dag\}_{qk}\hat\rho_{kq}\right\}} \\
\end{pmatrix} + 
\bcalM^{-\frac{1}{2}} 
\begin{pmatrix} 
 - \bOmega_g \bphi \\
   \bOmega_g^\ast \bphi^\ast \\
 - \C^{\left\{\frac{1}{2}\sum_{k,s,q,l=1}^M \{g^\dag\}_{ksql}\hat\rho_{kslq}\right\}} \\
 (\C^\ast)^{\left\{\frac{1}{2}\sum_{k,s,q,l=1}^M \{g^\dag\}_{lqsk}\hat\rho_{kslq}\right\}} \\
\end{pmatrix},
\eeq
where $\bphi=\{|\phi_k\rangle\}, k=1,\ldots,M$ is a column vector
and
$\bOmega_g = \{\Omega_{g,kq}\} 
= \left\{\sum^M_{s,l=1}\rho_{kslq}\{\hat{g}^\dag\}_{sl}\right\}$
a square matrix of dimension $M$.

With these ingredients at hand 
and inserting $\bcalP$ to the right 
(making use of the orthogonality of the response, Eq.~(\ref{LR_P_vector}), 
in the linear-response system or, equivalently,
that initially 
$\bcalP$ does not multiply the $\omega$ term; 
see Ref.~\cite{LR_MCTDHB}),
Eqs.~(\ref{LR_1st_orb_6}), (\ref{LR_1st_orb_7}),
(\ref{LR_1st_coeff_2}), and (\ref{LR_1st_coeff_3})
can be cast into the form:
\beq\label{LR_matrix_EO1}
\bcalM^{+\frac{1}{2}} 
\left(\left\{\bcalP \bcalM^{-\frac{1}{2}}\bcalL\bcalM^{-\frac{1}{2}}\bcalP\right\}
- \omega\right)
\left\{\bcalM^{+\frac{1}{2}} 
\begin{pmatrix} 
 \u \\
 \v \\
 \C_u \\
 \C_v \\
\end{pmatrix}\right\} =
\bcalM^{+\frac{1}{2}} \left\{\bcalP \bcalM^{+\frac{1}{2}}\bcalR\right\}.
\eeq
Multiply from the left by $\bcalM^{-\frac{1}{2}}$
and (to avoid cumbersome notation)
using the assignments:
\beq\label{LR_vector_assigments}
\left\{\bcalP \bcalM^{-\frac{1}{2}}\bcalL\bcalM^{-\frac{1}{2}}\bcalP\right\}
\Longrightarrow \bcalL, \quad
 \left\{\bcalP \bcalM^{+\frac{1}{2}}\bcalR\right\} \Longrightarrow \bcalR, \quad
\left\{\bcalM^{+\frac{1}{2}} 
\begin{pmatrix} 
 \u \\
 \v \\
 \C_u \\
 \C_v \\
\end{pmatrix}\right\} \Longrightarrow 
\begin{pmatrix} 
 \u \\
 \v \\
 \C_u \\
 \C_v \\
\end{pmatrix},
\eeq 
we arrive at the linear-response 
working matrix equation (\ref{LR_matrix_EO_working}).

%%%%%%%%%%%%%%%%%%%%%%%%%%%%%%%%%%%%%%%%%%%%%%%%%
\subsubsection{Solving the time-dependent 
identical-particle Schr\"odinger equation in linear response}\label{matrix_form_sec_s2}
%%%%%%%%%%%%%%%%%%%%%%%%%%%%%%%%%%%%%%%%%%%%%%%%%

We assume $\bcalL$ is diagonalizable 
and that all its eigenvalues are real 
(we have found this numerically 
to be the case  
for trapped repulsive bosons
in their ground state \cite{LR_MCTDHB}).
Conversely and physically, 
when not all eigenvalues are real,
an initial infinitesimal perturbation will grow up
(within linear response) exponentially,
implying that the system is unstable. 
We recall that, 
due to the projector matrix $\bcalP$
within $\bcalL$, 
the eigenvectors $\{\R^k\}$ are
in the complementary space, 
except for the zero excitations,
assuming the case that, generally, 
the many-particle ground state is not degenerate.

Before we proceed and use the eigenvectors
of $\bcalL$, Eq.~(\ref{LR_matrix_diag_1}), 
to solve 
the linear-response system Eq.~(\ref{LR_matrix_EO_working}),
we briefly discuss the symmetries
and other properties of the spectrum 
Eq.~(\ref{LR_matrix_diag_1}).
The additional 
ingredient we need in order to analyze the solutions of the 
linear-response matrix $\bcalL$ 
are the `spin' matrices:
\beq\label{LR_matrix_Sig1}
 {\bSig}_\1 =
\begin{pmatrix} 
 {\bSig}_{\1 }^o & {\0 }_{oc} \\
 {\0 }_{co} & {\bSig}_{\1 }^c \\
\end{pmatrix}, \qquad \qquad
 {\bSig}_{\1 }^o =
\begin{pmatrix} 
  {\O }_o & {\1 }_o \\
  {\1 }_o & {\O }_o \\
\end{pmatrix}, \quad
 {\bSig}_{\1 }^c =
\begin{pmatrix} 
  {\O }_c & {\1 }_c \\
  {\1 }_c & {\O }_c \\
\end{pmatrix}
\eeq
and
\beq\label{LR_matrix_Sig3}
 {\bSig}_\3 =
\begin{pmatrix} 
 {\bSig}_{\3 }^o & {\0 }_{oc} \\
 {\0 }_{co} & {\bSig}_{\3 }^c \\
\end{pmatrix}, \qquad \qquad
 {\bSig}_{\3 }^o =
\begin{pmatrix} 
  {\1 }_o & {\0 }_o \\
  {\0 }_o & {-\1 }_o \\
\end{pmatrix}, \quad
 {\bSig}_{\3 }^c =
\begin{pmatrix} 
  {\1 }_c & {\0 }_c \\
  {\0 }_c & {-\1 }_c \\
\end{pmatrix}.
\eeq

Making use of the ${\bSig}_\1 $ matrix, Eq.~(\ref{LR_matrix_Sig1}),
and examining each of the blocks of $\bcalL$, 
we find the symmetry
property:
\beq\label{LR_matrix_Sig1_symmetry}
{\bSig}_\1 \bcalL {\bSig}_\1 = -(\bcalL)^\ast.
\eeq
Similarly,
with the help of the ${\bSig}_\3 $ matrix, Eq.~(\ref{LR_matrix_Sig3}), 
we find \footnote{Of course, the metric $\bcalM$ and projector $\bcalP$ matrices
within the response matrix $\bcalL$ [see Eq.~(\ref{LR_vector_assigments})] 
comply with these symmetries, explicitly,
${\bSig}_\1 \bcalM {\bSig}_\1 = (\bcalM)^\ast$, 
${\bSig}_\1 \bcalP {\bSig}_\1 = (\bcalP)^\ast$
and
${\bSig}_\3 \bcalM {\bSig}_\3 = (\bcalM)^\dag$,
${\bSig}_\3 \bcalP {\bSig}_\3 = (\bcalP)^\dag$.}:
\beq\label{LR_matrix_Sig3_symmetry}
{\bSig}_\3 \bcalL {\bSig}_\3 = (\bcalL)^\dag.
\eeq
The symmetries Eqs.~(\ref{LR_matrix_Sig1_symmetry}) 
and (\ref{LR_matrix_Sig3_symmetry}) 
lead to the following properties of the spectrum $\{\omega^k\}$ and eigenvectors $\{\R^k\}$:
From Eq.~(\ref{LR_matrix_Sig1_symmetry}) we learn that
$\R^{-k} \equiv {\bSig}_\1 (\R^k)^\ast$ is an eigenvector of $\bcalL$
with the eigenvalue $-(\omega_k)^\ast$,
and from Eq.~(\ref{LR_matrix_Sig3_symmetry}) we can construct the
adjoint (or, left) eigenvectors 
$(\L^k)^\dag \bcalL = \omega^k (\L^k)^\dag$,
where $\L^k = \mathrm{sng}^k {\bSig}_\3 \R^k$
and $\mathrm{sng}^k$ stands for the sign of the `scalar product' $\{(\R^k)^\dag {\bSig}_\3 \R^k\}$.
The quantity $\mathrm{sng}^k$ is formally introduced 
because the `scalar product'
is with the metric ${\bSig}_\3 $ and thus can have a negative value,
when the contribution from the $v$ terms is larger than the contribution from the $u$ ones.
These allow us to obtain the orthogonality relations for eigenvectors with
excitation index $k$ and $k'$:
\beqn\label{left_right_orthogonality}
& &  (\L^k)^\dag \R^{k'} = \mathrm{sng}^k \left[ 
(\u^k)^\dag \u^{k'} - (\v^k)^\dag \v^{k'} + (\C_u^k)^\dag \C_u^{k'} - (\C_v^k)^\dag \C_v^{k'}
\right] = \delta_{kk'}, \\
& &  
(\L^k)^\dag \R^{-k'} =
\mathrm{sng}^k \left[ 
(\u^k)^\dag (\v^{k'})^\ast - (\v^k)^\dag (\u^{k'})^\ast + (\C_u^k)^\dag (\C_v^{k'})^\ast - (\C_v^k)^\dag (\C_u^{k'})^\ast
\right] = 0. \nonumber \
\eeqn

Finally, we can use these relations to write the resolution of the
identity $\1 $ and 
the spectral resolution of $\bcalL$ from the positive (non-negative) sector of $\{\R^k\}$.
Thus we have:
\beq\label{LR_identity_resolution}
 \1 = \sum_{k=0}^{M^2}  \left\{ \R_0^{k}  (\L_0^{k})^\dag +  \R_0^{-k} (\L_0^{-k})^\dag \right\}
+ \sum_{k>M^2} \left\{ \R^{k} (\L^{k})^\dag + \R^{-k} (\L^{-k})^\dag \right\},
\eeq
where
$\L^{-k} = - \mathrm{sng}^k {\bSig}_\3 \R^{-k} = 
- \mathrm{sng}^k {\bSig}_\3 {\bSig}_\1 (\R^{k})^\ast 
= {\bSig}_\1 (\L^k)^\ast$.
Note in comparison with $\L^{k}$ the minus sign in $\L^{-k}$,
which emerges from the fact that ${\bSig}_\3 $ and ${\bSig}_\1 $ anti-commute.
In Eq.~(\ref{LR_identity_resolution}) 
the first group of vectors 
are the zero-mode excitations.
There are $2(M^2+1)$ such eigenvectors.
$M^2$ eigenvectors are obtained by putting any of the $M$ ground-state orbitals $|\phi_q\rangle, q=1,\ldots,M$ 
in any of the $M$ entries of the $\u^k$ vector and 1 eigenvector is obtained 
by taking the ground-state vector of coefficients $\C$ as the vector $\C_u$.
The other entries of $\R_0^{k}$ are all zero.
This amount doubles on the account of ${\bSig}_\1 $ and
the negative (non-positive) sector.
The second group of vectors in Eq.~(\ref{LR_identity_resolution})
are the non-zero-mode excitations.
The respective summation index, $k > M^2$,
indicates that we
enumerate them after the group of zero-mode excitations
for which the index of enumeration satisfies $k \in [0,M^2]$.
Finally, for $\bcalL$ we have:
\beq\label{LR_calL_matrix_resolution}
 \bcalL = \sum_{k>M^2} \omega_k \left\{ \R^{k} (\L^{k})^\dag  -
              \R^{-k} (\L^{-k})^\dag \right\}.
\eeq
It is instructive to compare the structure of Eqs.~(\ref{LR_identity_resolution}) and (\ref{LR_calL_matrix_resolution})
to their analogs derived from the full Schr\"odinger equation, 
Eqs.~(\ref{SE_12_LR_identity})
and (\ref{SE_13_LR_L_matrix}).

We now return to Eq.~(\ref{LR_matrix_EO_working}) and employ
the eigenvectors of $\bcalL$ to solve it.
To this end we expand the response amplitudes and perturbation as follows:
\beq\label{LR_expand_k_1}
\begin{pmatrix} 
 \u \\
 \v \\
 \C_u \\
 \C_v \\
\end{pmatrix} = \sum_{k} c_k \R^{k} =
\sum_{k>M^2} [c_k \R^{k} + c_{-k} \R^{-k}], \qquad
\bcalR =  - \sum_{k} \gamma_k \R^{k} =  - \sum_{k>M^2} [\gamma_k \R^{k} + \gamma_{-k} \R^{-k}],
\eeq 
where, to remind, $\R^{-k} \equiv {\bSig}_\1 (\R^k)^\ast$.
The zero-mode eigenvectors, as discussed above,
do not contribute.
Substituting Eq.~(\ref{LR_expand_k_1}) 
into Eq.~(\ref{LR_matrix_EO_working})
we find
\beq\label{LR_expand_k_2}
\begin{pmatrix} 
 \u \\
 \v \\
 \C_u \\
 \C_v \\
\end{pmatrix} = \sum_{k>M^2} \left[\frac{\gamma_k}{\omega-\omega_k} \R^{k} + \frac{\gamma_{-k}}{\omega+\omega_{k}} \R^{-k}\right], 
\eeq
where the response weights are given explicitly by [see Eqs.~(\ref{LR_perturbing_fields}) and (\ref{LR_vector_assigments})], $k>M^2$:
\beqn\label{response_wg_MCTDHX}
& &  \gamma_k = - (\L^{k})^\dag \bcalR = 
\mathrm{sng}^k \Bigg \{ (\u^k)^\dag \left[\brho^{+\frac{1}{2}} \hat f^\dag + \brho^{-\frac{1}{2}}\bOmega_g\right] \bphi +
  (\v^k)^\dag  \left[\brho^{+\frac{1}{2}} \hat f^\ast + \brho^{-\frac{1}{2}} \bOmega_g^\ast\right] \bphi^\ast + \nonumber \\
& & \qquad + (\C_u^k)^\dag \cdot 
\C^{\left\{\sum_{k,q=1}^M \{f^\dag\}_{kq}\hat\rho_{kq} + \frac{1}{2}\sum_{k,s,q,l=1}^M \{g^\dag\}_{ksql}\hat\rho_{kslq}\right\}} + \nonumber \\
& &   \qquad   +  (\C_v^k)^\dag \cdot 
(\C^\ast)^{\left\{\sum_{k,q=1}^M \{f^\dag\}_{qk}\hat\rho_{kq} + \frac{1}{2}\sum_{k,s,q,l=1}^M \{g^\dag\}_{lqsk}\hat\rho_{kslq}\right\}} \Bigg \},
\nonumber \\
& &  \gamma_{-k} = - (\L^{-k})^\dag \bcalR = 
- \mathrm{sng}^k \Bigg \{ (\v^k)^t \left[\brho^{+\frac{1}{2}} \hat f^\dag + \brho^{-\frac{1}{2}}\bOmega_g\right] \bphi +
  (\u^k)^t \left[\brho^{+\frac{1}{2}} \hat f^\ast + \brho^{-\frac{1}{2}} \bOmega_g^\ast\right] \bphi^\ast + \nonumber \\
& & \qquad + (\C_v^k)^t \cdot 
\C^{\left\{\sum_{k,q=1}^M \{f^\dag\}_{kq}\hat\rho_{kq} + \frac{1}{2}\sum_{k,s,q,l=1}^M \{g^\dag\}_{ksql}\hat\rho_{kslq}\right\}} + \nonumber \\
& &   \qquad   +  (\C_u^k)^t \cdot 
(\C^\ast)^{\left\{\sum_{k,q=1}^M \{f^\dag\}_{qk}\hat\rho_{kq} + \frac{1}{2}\sum_{k,s,q,l=1}^M \{g^\dag\}_{lqsk}\hat\rho_{kslq}\right\}} \Bigg \}. 
\eeqn
From the right-hand sides of Eq.~(\ref{response_wg_MCTDHX}) we see that the response weights combine 
the
contributions from the responses of all orbitals and of all expansion coefficients.
Note that $\bcalP$, see Eq.~(\ref{LR_vector_assigments}), 
falls out of the expression 
for the response weights.  

Reinserting the expansion for the response amplitudes, Eq.~(\ref{LR_expand_k_2}), 
into the ansatz for the orbitals and expansion coefficients, Eq.~(\ref{LR_ansatz}), gives
their time dependence in linear response:
\beqn\label{orbitals_coeff_final_res_MCTDHX}
& & \bphi(\r,t) \approx \bphi(\r) + \delta\bphi(\r,t), \nonumber \\
& & \qquad \delta\bphi(\r,t) = \sum_{k>M^2} 
 \Bigg\{ \left[\gamma_k  \{\brho\}^{-\frac{1}{2}} \u^k(\r) e^{-i \omega t} + 
       \gamma_k^\ast \{\brho^\ast\}^{-\frac{1}{2}}  \{\v^k(\r)\}^\ast e^{+i \omega t} \right]/(\omega-\omega_k) + \nonumber \\
& & \qquad \qquad + \left[\gamma_{-k}  \{\brho\}^{-\frac{1}{2}}  \{\v^k(\r)\}^\ast e^{-i \omega t} + 
       \gamma_{-k}^\ast \{\brho^\ast\}^{-\frac{1}{2}}  \u^k(\r) e^{+i \omega t} \right]/(\omega+\omega_k) \Bigg\}, \nonumber \\
& & \C(t) \approx \C + \delta\C(t), \nonumber \\
& & \qquad \delta\C(t) = \sum_{k>M^2} 
\Bigg\{ \left[\gamma_k \C_u^k e^{-i \omega t} + 
       \gamma_k^\ast  \{\C_v^k\}^\ast e^{+i \omega t} \right]/(\omega-\omega_k) + \nonumber \\
& & \qquad \qquad + \left[\gamma_{-k}  \{\C_v^k\}^\ast  e^{-i \omega t} + 
       \gamma_{-k}^\ast  \C_u^k e^{+i \omega t} \right]/(\omega+\omega_k) \Bigg\},
\eeqn
with $\delta\bphi(\r,t) = \{\delta\phi_j(\r,t)\}, j=1,\ldots, M$.
Thus, the orbitals and the expansion coefficients show the largest
response at the frequencies $\{\pm \omega_k\}$. 
Moreover, the response at a given frequency $\omega_k$ 
is not necessarily equally strong for all
the orbitals $\bphi(\r,t)$ and similarly for all the expansion coefficients $\C(t)$.
The reason is because the components of the response
amplitudes $\u^k$, $\v^k$, $\C_u^k$, and $\C_v^k$
are not individually normalized, 
but rather the whole amplitude vector is,
see Eq.~(\ref{left_right_orthogonality}).

Finally, from Eq.~(\ref{orbitals_coeff_final_res_MCTDHX})
the time-dependent many-particle wavefunction is given in 
linear response by:
\beqn
& & |\Psi(t)\rangle \approx \sum_{\vec n} C_{\vec n} |\vec n\rangle +   \sum_{\vec n} \delta C_{\vec n}(t) |\vec n\rangle +  \\
& & \quad +  \sum_{\vec n} C_{\vec n} \left[ \sum_{j=1}^{M} 
(\pm 1)^{\{\sum_{l=j+1}^M n_l\}} \sqrt{n_j} \sqrt{\langle \delta\phi_j(\r,t)|\delta\phi_j(\r,t)\rangle}
|n_1,\ldots,n_j-1,\ldots,n_M,1_{M+1}^j;t\rangle \right] \nonumber,
\eeqn
where $|1_{M+1}^j;t\rangle$ is associated with the (unnormalized) time-dependent response orbital $\delta\phi_j(\r,t)$.
The response orbitals $\{\delta\phi_j(\r,t)\}$ need not be orthogonal to each other,
unlike their orthogonality with the ground-state orbitals which originates from
the orbital 
differential condition Eq.~(\ref{diff_cond}).
This concludes our derivation of LR-MCTDHB and LR-MCTDHF,
in a unified manner and representation.

%%%%%%%%%%%%%%%%%%%%%%%%%%%%%%%%%%%%%%%%%%%%%%%%
\section{Linear response in the multiconfigurational time-dependent Hartree
framework for distinguishable degrees-of-freedom}\label{dis_sec}
%%%%%%%%%%%%%%%%%%%%%%%%%%%%%%%%%%%%%%%%%%%%%%%%

This section deals with systems of distinguishable degrees-of-freedom and -- starting from
the MCTDH propagation theory -- it develops
the corresponding linear-response theory, which we denote by LR-MCTDH,
based on the knowhow of the previous section, Sec.~\ref{indis_sec},
and some new ingredients.

%%%%%%%%%%%%%%%%%%%%%%%%%%%%%%%%%%%%%%%%%%%%%%%%
\subsection{Quick derivation of MCTDH: Basic and new ingredients and notations}
%%%%%%%%%%%%%%%%%%%%%%%%%%%%%%%%%%%%%%%%%%%%%%%%

Let us have $j=1,\ldots,Q$ in-general distinguishable degrees-of-freedom which
we label by the generalized coordinates $\r_1,\ldots,\r_Q$.
Each degree-of-freedom is expanded 
by $n_j=1,\ldots,M_j$ 
orthonormal time-dependent orbitals.
Within a concise (yet clear) notation,
the many-particle wavefunction takes on the 
following appearance: 
\beq\label{Psi_MCTDH}
|\Psi(t)\rangle = \sum_{\vec n} C_{\vec n} |\vec n;t\rangle = 
\sum_{\vec n[j]}  \sum_{n_j=1}^{M_j} 
C_{\vec n[j] n_j} |\vec n[j];t\rangle |n_j;t\rangle, \quad \forall j.
\eeq
The left-hand side of Eq.~(\ref{Psi_MCTDH}) looks the same as for identical particles, 
but the meaning is, of course, different.
Namely, the $|\vec n;t\rangle$ are configurations--Hartree products,
$|\vec n;t\rangle = \prod_{j=1}^Q |n_j;t\rangle$, 
not permanents or determinants.

The generic Hamiltonian of the $Q$ coupled degrees-of-freedom is:
\beq\label{Ham_MCTDH}
 \hat H = \sum_{j=1}^Q \hat h^j(\r_j) + \hat W(\r_1,\ldots,\r_Q).
\eeq
The coupling--interaction part is written generically.
For instance, it can be comprised of few-body ingredients
or, in the general case, couple all $Q$ degrees-of-freedom.

The action--functional reads (time-dependence of quantities 
is suppressed when unambiguous):
\beq\label{action_MCTDH}
 S = \int dt \left\{ \langle\Psi|\hat H - i\frac{\partial}{\partial t}|\Psi\rangle -
 \sum_{j=1}^Q \sum_{n_j,m_j}^{M_j} 
\mu_{n_jm_j}^j (\langle n_j|m_j\rangle - \delta_{n_jm_j}) 
-\varepsilon \C^\dag\C \right\}.
\eeq
The MCTDH equations of motion for the time-dependent 
orbitals and expansion coefficients \cite{MCTDH_cpl,MCTDH_jcp} are readily  
derived by equating the variation of $S$ with respect to these quantities to zero.
Thus one finds: 
\beqn\label{MCTDH_equ}
& &  \hat {\mathbf P}_j \sum_{m_j=1}^{M_j} [\rho^j_{n_jm_j} \hat h^j 
+ \hat \Omega^j_{n_jm_j}] |m_j\rangle = 
i \sum_{m_j=1}^{M_j} \rho^j_{n_jm_j} |\dot m_j\rangle, 
\qquad n_j=1,\ldots,M_j, \ j=1,\ldots,Q, \nonumber \\
& & \qquad {\mathbf H}(t)\C(t) = i\frac{\partial \C(t)}{\partial t}, \qquad
 H_{\vec{n}\vec{n}'}(t) = \langle\vec{n};t|\hat H|\vec{n}';t\rangle, \
\eeqn
where for each of the $j=1,\ldots,Q$ degrees-of-freedom we define the reduced one-body
density matrix:
\beq\label{MCTDH_r1}
 \rho^j_{n_jm_j} = \sum_{\vec n[j]} C^\ast_{\vec n[j] n_j} C_{\vec n[j] m_j}, 
\qquad n_j,m_j = 1,\ldots,M_j,
\eeq
the mean-field operators:
\beq\label{MCTDH_MFs}
 \hat \Omega^j_{n_jm_j} = \sum_{\vec n[j],\vec m[j]} \rho_{\vec n \vec m}
 \hat W_{\vec n[j] \vec m[j]}, \qquad n_j,m_j = 1,\ldots,M_j,
\eeq
with $\hat W_{\vec n[j] \vec m[j]}=\langle\vec n[j]|\hat W|\vec m[j]\rangle$, 
and $\rho_{\vec n \vec m} = C^\ast_{\vec n}C_{\vec m} =
\rho_{\vec n[j]n_j \vec m[j]m_j} = C^\ast_{\vec n[j] n_j}C_{\vec m[j] m_j}, \forall j$
is the (reduced) all-body density matrix.
Finally, the projectors are defined as:
\beq\label{MCTDH_projectors}
\hat {\mathbf P}_j = 1 - \sum_{n'_j=1}^{M_j} |n'_j\rangle\langle n'_j|.  
\eeq

We now proceed in the same manner as done in Sec.~\ref{indis_sec} 
above with the coefficients.
The ``fully projected" representation of MCTDH is obtained by
assigning the coefficients as follows 
[see Eq.~(\ref{P_MCTDHX_phase})]: 
$\C \rightarrow \C e^{-i \int^t dt' \C^\dag(t') \H(t') \C(t')}$.
This results in:
\beqn\label{P_MCTDH_equ}
& &  \hat {\mathbf P}_j \sum_{m_j=1}^{M_j} [\rho^j_{n_jm_j} \hat h^j + \hat \Omega^j_{n_jm_j}] |m_j\rangle = 
i \sum_{m_j=1}^{M_j} \rho^j_{n_jm_j} |\dot m_j\rangle, \qquad n_j=1,\ldots,M_j, \ j=1,\ldots,Q, \nonumber \\
& & \qquad \qquad {\mathbf P}_C(t) {\mathbf H}(t)\C(t) = i\frac{\partial \C(t)}{\partial t}, \
\eeqn
with ${\mathbf P}_C = \1 - \C\C^\dag$.
In Eq.~(\ref{P_MCTDH_equ}) both
the orbitals \cite{MCTDH_cpl,MCTDH_jcp} and the expansion coefficients satisfy
the differential conditions:
\beq\label{diff_cond_MCTDH_both}
 i\langle n_j|\dot m_j\rangle = 0, \quad n_j,m_j=1,\ldots,M_j, \, \, j=1,\ldots,Q, \qquad \qquad  i\C^\dag\dot\C = 0,
\eeq
namely the evolution of the system's wavefunction, Eq.~(\ref{Psi_MCTDH}), 
is completely orthogonal:
$i\langle\Psi(t)|\dot\Psi(t)\rangle = 0$.

Finally, from either Eq.~(\ref{MCTDH_equ}) or Eq.~(\ref{P_MCTDH_equ}) 
the time-independent MCH theory is obtained:
\beqn\label{MCH_equ}
& & \hat {\mathbf P}_j \sum_{m_j=1}^{M_j} [\rho^j_{n_jm_j} \hat h^j + \hat \Omega^j_{n_jm_j}] |m_j\rangle = 0, 
\qquad n_j=1,\ldots,M_j, \ j=1,\ldots,Q, \qquad  \Longleftrightarrow \nonumber \\
& & \sum_{m_j=1}^{M_j} [\rho^j_{n_jm_j} \hat h^j - \mu_{n_jm_j}^j + \hat \Omega^j_{n_jm_j}] |m_j\rangle = 0, 
\qquad n_j=1,\ldots,M_j, \ j=1,\ldots,Q, \nonumber \\
& & \mu_{n_jm_j}^j = \langle m_j|\sum_{n'_j=1}^{M_j} [\rho^j_{n_jn'_j} \hat h^j + \hat \Omega^j_{n_jn'_j}] |n'_j\rangle = 0, 
\qquad n_j,m_j=1,\ldots,M_j, \ j=1,\ldots,Q, \nonumber \\
& &  {\mathbf P}_C {\mathbf H}\C = \0 
\qquad \Longleftrightarrow \qquad {\mathbf H}\C =\varepsilon\C, \qquad
 H_{\vec{n}\vec{n}'} = \langle\vec{n}|\hat H|\vec{n}'\rangle. \
\eeqn

Based on the linear response
for indistinguishable particles of Sec.~\ref{indis_sec} 
and the above notation for MCTDH, 
we now proceed to derive LR-MCTDH.
Only the essential formulas will be presented 
in detail.

%%%%%%%%%%%%%%%%%%%%%%%%%%%%%%%%%%%%%%%%%%%%%%%%
\subsection{Perturbation and variation}
%%%%%%%%%%%%%%%%%%%%%%%%%%%%%%%%%%%%%%%%%%%%%%%%

We derive the linear response of MCTDH using a small
perturbation {\it around} the MCH solution, typically the ground state.
We will consider both a one-body time-dependent perturbation, 
say an external field,
and a generic ``all-body" time-dependent perturbation, 
e.g., a change to the potential-energy hyper-surface between all degrees-of-freedom.
The respective ansatz is as follows:
\beqn\label{MCTDH_LR_ansatz}
& & |n_j;t\rangle \approx |n_j\rangle + |\delta n_j;t\rangle, 
\qquad n_j=1,\ldots,M_j, \ j=1,\ldots,Q, \nonumber \\
& & |\delta n_j;t\rangle = 
|u_{n_j}\rangle e^{-i \omega t} + |v^\ast_{n_j}\rangle e^{+i \omega t},  \qquad n_j=1,\ldots,M_j, \ j=1,\ldots,Q, \nonumber \\
& & \C(t) \approx [\C + \delta\C(t)]
\quad \Longleftrightarrow \quad 
\C(t) \approx e^{-i \varepsilon t} [\C + \delta\C(t)]
\quad \mathrm{(without \ coefficients' \ projector)}, \nonumber \\
& & \delta\C(t) = \C_u e^{-i \omega t} + \C^\ast_v e^{+i \omega t}, \nonumber \\
& & \delta \hat h^j(\r_j,t) = \hat f_j^\dag(\r_j) e^{-i \omega t} + \hat f_j(\r_j) e^{+i \omega t}, \qquad j=1,\ldots,Q, \nonumber \\
& & \delta \hat W(\r_1,\ldots,\r_Q,t) = \hat g^\dag(\r_1,\ldots,\r_Q) e^{-i \omega t} + \hat g(\r_1,\ldots,\r_Q) e^{+i \omega t}. \
\eeqn
Here, $\{|\delta n_j;t\rangle\}, \, j=1,\ldots,Q$ and $\delta\C(t)$
are the perturbed parts of the orbitals and
coefficients, respectively.
The perturbed parts of the system's wavefunction are
all comprised of $u$ and $v$ contributions.
The operators $\{\hat f_j(\r)\}, \, j=1,\ldots,Q$ and $\hat g(\r_1,\ldots,\r_Q)$ generate 
independently the one-body and all-body perturbations.
It is instructive to contrast the perturbation ansatz Eq.~(\ref{MCTDH_LR_ansatz}) 
for distinguishable particles and the respective one for identical particles, Eq.~(\ref{LR_ansatz}).
In particular, {\it each} of the $j=1,\ldots,Q$ distinct degrees-of-freedom
can, in principle, be perturbed by a {\it different} one-body operator, $\hat f_j$.
We emphasize that despite the distinguishability,
the whole coupled system responds as a whole,
even if only, say the $j_0$-th degree-of-freedom is perturbed,
and the remaining ones are not, 
namely $\hat f_{j \ne j_0} = 0$ and $\hat g = 0$.
As was used above, 
the perturbing frequency $\omega$ is assumed to be non-zero.
The Hamiltonian including the perturbation is hermitian.

In a similar manner like in Sec.~\ref{indis_sec} 
one can derive the orthogonality conditions between the perturbations
and ground-state quantities (orbitals and coefficients).
We will not repeat this step here.

%%%%%%%%%%%%%%%%%%%%%%%%%%%%%%%%%%%%%%%%%%%%%%%%
\subsection{The linear-response system and its formal solution}\label{MCTDH_sub_LR_eqs}
%%%%%%%%%%%%%%%%%%%%%%%%%%%%%%%%%%%%%%%%%%%%%%%%

Our goal now is to develop the linear-response theory and solve the 
respective many-body Schr\"odinger equation in linear response.
To assist the reader,
we have divided the task to three.
First, in Subsec.~\ref{MCTDH_sub_LR_eqs_s1}
we linearize the MCTDH theory and derive 
separately the orbitals' and coefficients' linear-response equations.
Then, in Subsec.~\ref{MCTDH_sub_LR_eqs_s2},
we cast them into a matrix form, 
adapting the same strategy as done for identical particles above.
Finally,  in Subsec.~\ref{MCTDH_sub_LR_eqs_s3}, 
after discussing the symmetries and other 
properties of the
linear-response matrix system we solve for the perturbed 
time-dependent orbitals and coefficients,
and the MCTDH wavefunction 
in linear response.

%%%%%%%%%%%%%%%%%%%%%%%%%%%%%%%%%%%%%%%%%%%%%%%%
\subsubsection{Linear-response equations}\label{MCTDH_sub_LR_eqs_s1}
%%%%%%%%%%%%%%%%%%%%%%%%%%%%%%%%%%%%%%%%%%%%%%%%

Substituting the ansatz Eq.~(\ref{MCTDH_LR_ansatz})
 into the MCTDH equations of motion,
either into the standard form, Eq.~(\ref{MCTDH_equ}), or into the ``fully projected" form, Eq.~(\ref{P_MCTDH_equ}), 
we find to 0-th order the MCH equations themselves, see Eq.~(\ref{MCH_equ}).
For the linear-response (1-st order) equations we will proceed, as above,
separately for the orbitals and for the coefficients.
To derive the equations for the coefficients, 
we will work with the ``fully projected" form, Eq.~(\ref{P_MCTDH_equ}),
namely employ explicitly the coefficients' projector ${\mathbf P}_C$.

We find from either the standard, Eq.~(\ref{MCTDH_equ}), or the 
``fully projected" form, Eq.~(\ref{P_MCTDH_equ}), 
of MCTDH
the following relation for the perturbed orbitals:
\beqn\label{LR_MCTDH_orbitals_1}
& &  \hat {\mathbf P}_j \sum_{m_j=1}^{M_j} [\delta\rho^j_{n_jm_j} \hat h^j + \rho^j_{n_jm_j} \delta\hat h^j 
+ \delta\hat \Omega^j_{n_jm_j}] |m_j\rangle + \nonumber \\
& & + \hat {\mathbf P}_j \sum_{m_j=1}^{M_j} [\rho^j_{n_jm_j} \hat h^j - \mu^j_{n_jm_j} + \hat \Omega^j_{n_jm_j}] |\delta m_j\rangle = \nonumber \\
& & = i \sum_{m_j=1}^{M_j} \rho^j_{n_jm_j} |\delta \dot m_j\rangle, \qquad n_j=1,\ldots,M_j, \ j=1,\ldots,Q, \
\eeqn
where we have used 
$\delta \hat {\mathbf P}_j \sum_{m_j=1}^{M_j} [\rho^j_{n_jm_j} \hat h^j + \hat \Omega^j_{n_jm_j}] |m_j\rangle = 
- \hat {\mathbf P}_j \sum_{m_j=1}^{M_j} \mu^j_{n_jm_j} |\delta m_j\rangle,  n_j=1,\ldots,M_j, \ j=1,\ldots,Q$.
Eq.~(\ref{LR_MCTDH_orbitals_1}) is the generic form for 
the linear-response equations
for the orbitals' part of $Q$ 
coupled degrees-of-freedom.
In the same manner that for $M_1=\ldots=M_Q=1$ MCTDH boils down to the text-book
time-dependent Hartree (TDH) equations, see, e.g., Ref.~\cite{MCTDH_review},
Eq.~(\ref{LR_MCTDH_orbitals_1}) boils down to the linear response of TDH
(LR-TDH), 
which can readily be prescribed; 
we do not write explicitly the resulting 
equations 
here.

Let us proceed to 
the main steps in the 
derivation of the LR-MCTDH equations.
We need to consider the variation leading to the $e^{\mp i \omega t}$ terms separately.
This leads to
the following ingredients for 
the $e^{-i \omega t}$ equation:
\beqn\label{LR_MCTDH_var_E1}
 \hat W_{\vec n[j] \vec m[j]}  & &  \Longrightarrow \nonumber \\
  \delta{\hat W_{\vec n[j] \vec m[j]}}|_{e^{- i\omega t}} = & &
\sum^{Q}_{k\ne j=1}
(\hat W_{\vec n[j,k]\vec m[j]v_{n_k}} + 
 \hat W_{\vec n[j] \vec m[j,k] u_{m_k}}), \nonumber \\
  \hat \Omega^j_{n_jm_j} & & 
 \Longrightarrow \nonumber \\
 {\delta\hat \Omega^j_{n_jm_j}}|_{e^{- i\omega t}} = & &
\sum_{\vec n[j],\vec m[j]} (\{\C^t \brho_{\vec m \vec n} \cdot \C_v + 
\C^\dag \brho_{\vec n \vec m} \cdot \C_u\} \hat W_{\vec n[j] \vec m[j]} +
 \rho_{\vec n \vec m}
\delta{\hat W_{\vec n[j] \vec m[j]}}|_{e^{- i\omega t}}), \nonumber \\
 \rho_{n_jm_j}^j 
& & \Longrightarrow \nonumber \\ 
 {\delta\rho_{n_jm_j}^j}|_{e^{- i\omega t}} = & & 
  \C^t \brho_{m_jn_j}^j \cdot \C_v +  \C^\dag \brho_{n_jm_j}^j \cdot \C_u,  \
\eeqn
where we have used a tensor-product representation
to work with the vector of coefficients and its
response--variation explicitly, 
see Appendix \ref{MCTDH_tensor_appendix}.
In the above equation 
for the variation of $\hat W_{\vec n[j] \vec m[j]}$
with the summation over
$k\ne j$ 
it is implicitly taken that
$n_k\in\vec n[j], m_k\in\vec m[j]$.

With the ingredients in Eq.~(\ref{LR_MCTDH_var_E1}) 
we get the explicit first-order equation
associated with the $e^{-i\omega t}$ term:
\beqn\label{LR_MCTDH_var_E2}
& &  \hat {\mathbf P}_j \sum_{m_j=1}^{M_j} 
[ \{\C^t \brho_{m_jn_j}^j \cdot \C_v +  \C^\dag \brho_{n_jm_j}^j \cdot \C_u\}
 \hat h^j + \nonumber \\
 & & +\sum_{\vec n[j],\vec m[j]} \{\C^t \brho_{\vec m \vec n} \cdot \C_v + 
\C^\dag \brho_{\vec n \vec m} \cdot \C_u\} \hat W_{\vec n[j] \vec m[j]}]
|m_j\rangle + \nonumber \\
& & +\hat {\mathbf P}_j \sum_{m_j=1}^{M_j} [\rho^j_{n_jm_j}
(\hat h^j - \omega) - \mu^j_{n_jm_j} + \hat \Omega^j_{n_jm_j}] |u_{m_j}\rangle +  \\
& & +\hat {\mathbf P}_j 
\sum^{Q}_{k\ne j=1} 
\sum_{m_k=1}^{M_k} \sum_{\vec n[j],\vec m[k]} 
\rho_{\vec n \vec m}
(\hat K_{\vec n[j]\vec m[k]}|u_{m_k}\rangle + 
 \hat K_{\vec n[j,k] \vec m}|v_{n_k}\rangle) 
= \nonumber \\
& & = - \hat {\mathbf P}_j \sum_{m_j=1}^{M_j} (\rho^j_{n_jm_j} \hat f^\dag_j +
\sum_{\vec n[j],\vec m[j]} \rho_{\vec n \vec m} 
\{\hat g^\dag\}_{\vec n[j] \vec m[j]}) |m_j\rangle, \qquad
n_j=1,\ldots,M_j, \ j=1,\ldots,Q. \nonumber \
\eeqn
Note the similarity and differences between Eq.~(\ref{LR_MCTDH_var_E2})
for distinguishable degrees-of-freedom
and Eq.~(\ref{LR_1st_orb_6}) for identical particles.
In particular, the responses of {\it different} degrees-of-freedom 
are coupled by the exchange-like potentials:
\beq\label{LR_MCTDH_exchange_like}
 \hat K_{\vec n[j]\vec m[k]}|u_{m_k}\rangle \equiv
 \hat W_{\vec n[j]\vec m[j,k]u_{m_k}}|m_j\rangle, \qquad
 \hat K_{\vec n[j,k]\vec m}|v_{n_k}\rangle \equiv
 \hat W_{\vec n[j,k]\vec m[j]v_{n_k}}|m_j\rangle.
\eeq
We will return to this point below.

Similarly, the final result for the first-order equation
associated with the $e^{+i\omega t}$ term is 
given by:
\beqn\label{LR_MCTDH_var_E3}
& &  \hat {\mathbf P}_j^\ast \sum_{m_j=1}^{M_j} 
[ \{\C^t \brho_{n_jm_j}^j \cdot \C_v +  \C^\dag \brho_{m_jn_j}^j \cdot \C_u\}
  \{\hat h^j\}^\ast + \nonumber \\
& & +\sum_{\vec n[j],\vec m[j]} \{\C^t \brho_{\vec n \vec m} \cdot \C_v + 
\C^\dag \brho_{\vec m \vec n} \cdot \C_u\} \hat W_{\vec m[j] \vec n[j]}]
|m_j^\ast\rangle + \nonumber \\
& & +\hat {\mathbf P}_j^\ast \sum_{m_j=1}^{M_j} [\rho^j_{m_jn_j}
(\{\hat h^j\}^\ast + \omega) - \mu^j_{m_jn_j} + 
\hat \Omega^j_{m_jn_j}] |v_{m_j}\rangle +  \\
& & + \hat {\mathbf P}_j^\ast 
\sum^{Q}_{k\ne j=1} 
\sum_{m_k=1}^{M_k} 
\sum_{\vec n[j],\vec m[k]} \rho_{\vec m \vec n}
(\hat K_{\vec m \vec n[j,k]} |u_{n_k}\rangle + 
\hat K_{\vec m[k] \vec n[j] } |v_{m_k}\rangle) 
= \nonumber \\
& & = - \hat {\mathbf P}_j^\ast \sum_{m_j=1}^{M_j} (\rho^j_{m_jn_j} \hat f^\ast_j +
\sum_{\vec n[j],\vec m[j]} \rho_{\vec m \vec n}
\{\hat g^\dag\}_{\vec m[j] \vec n[j]}) |m^\ast_j\rangle, \qquad
n_j=1,\ldots,M_j, \ j=1,\ldots,Q, \nonumber \
\eeqn
where
\beq\label{LR_MCTDH_exchange_like_2}
 \hat K_{\vec m \vec n[j,k]} |u_{n_k}\rangle \equiv
 \hat W_{\vec m[j]\vec n[j,k] u_{n_k}}|m^\ast_j\rangle, \qquad
 \hat K_{\vec m[k]\vec n[j] } |v_{m_k}\rangle \equiv
 \hat W_{\vec m[j,k]\vec n[j] v_{m_k}}|m^\ast_j\rangle.
\eeq
The connection between the $e^{\mp i\omega t}$ 
equations stems from (the complex conjugation and) the relations:
\beqn\label{LR_MCTDH_var_E4}
{[\delta\rho^j_{n_jm_j}|_{e^{+ i\omega t}}]}^\dag & & 
\qquad \Longleftrightarrow \qquad \delta\rho^j_{m_jn_j}|_{e^{- i\omega t}}, \nonumber \\
{[\delta\rho_{\vec n \vec m}|_{e^{+ i\omega t}}]}^\dag & & 
\qquad \Longleftrightarrow \qquad 
\delta\rho_{\vec m \vec n}|_{e^{- i\omega t}}, \nonumber \\
{[\delta\hat W_{\vec n[j] \vec m[j]}|_{e^{+ i\omega t}}]}^\dag & & 
\qquad \Longleftrightarrow \qquad \delta\hat W_{\vec m[j] \vec n[j]}|_{e^{- i\omega t}}, \nonumber \\
{[\delta\hat \Omega^j_{n_jm_j}|_{e^{+ i\omega t}}]}^\dag & & 
\qquad \Longleftrightarrow \qquad \delta\hat \Omega^j_{m_jn_j}|_{e^{- i\omega t}},  \
\eeqn
as for the 
identical-particle
case, see Eq.~(\ref{del_e_plus_1}).

How to boil down the LR-MCTDH equations when the all-body potential (and perturbations) 
are written as sums of products of one-body operators,
a useful representation within the MCTDH algorithm
\cite{MCTDH_book,MCTDH_Package},
is straightforward 
and will not be pursued here.

We now move to the perturbed coefficients,
namely to the linear-response equations in $1$-st order of the coefficients' part.
For the ``fully projected'' case  (we will only treat it
within LR-MCTDH) we find that:
\beq\label{coeff_LR_MCTDH_1}
 {\mathbf P}_C [(\H - \varepsilon)\delta\C + (\delta\H)\C] = i\delta\dot\C,
\eeq
just like Eq.~(\ref{P_LR_1st_coeff_1}) which utilized Eq.~(\ref{C_pro_inter_by_part}) 
in the indistinguishable-particle system.
Using a tensor-product representation (see Appendix \ref{MCTDH_tensor_appendix})
and in analogy to Eqs.~(\ref{Ham_den}) and (\ref{del_Ham_den}) 
for identical particles,
the Hamiltonian matrix:
\beq\label{n_Ham}
\H = \sum_{j=1}^{Q} \sum_{n_j,m_j=1}^{M_j} h^j_{n_jm_j} \brho_{n_jm_j}^j +
\sum_{\vec n,\vec m} W_{\vec n\vec m} \brho_{\vec n\vec m},
\eeq
and its variation:
\beqn\label{n_Ham_var}
\delta\H \!\! & & = \sum_{j=1}^{Q} \sum_{n_j,m_j=1}^{M_j} 
(h^j_{\delta n_jm_j} + h^j_{n_j \delta m_j} + \{\delta h^j\}_{n_jm_j})\brho_{n_jm_j}^j +
\sum_{\vec n,\vec m} 
(W_{\delta\vec n\vec m} + W_{\vec n\delta\vec m} + \{\delta W\}_{\vec n\vec m})
\brho_{\vec n\vec m}, \nonumber \\
& & = 
\sum_{j=1}^{Q} \sum_{n_j,m_j=1}^{M_j} 
[(\{{h^j}^\ast\}_{m_j^\ast \delta n_j^\ast} 
+ h^j_{n_j \delta m_j} + \{\delta h^j\}_{n_jm_j})\brho_{n_jm_j}^j + \nonumber \\
& & + \sum_{\vec n[j],\vec m[j]} 
(W_{\vec n[j]\vec m[j]m_j^\ast\delta n_j^\ast} + 
 W_{\vec n[j]\vec m[j]n_j\delta m_j})\brho_{\vec n\vec m}] + 
\sum_{\vec n,\vec m} \{\delta W\}_{\vec n\vec m} \brho_{\vec n\vec m} 
\eeqn
are found.
The first-order equation associated with the $e^{-i\omega t}$ term 
of the coefficients' response is then given by:
\beqn\label{coeff_LR_MCTDH_2}
& & {\mathbf P}_C (\omega + \varepsilon - \H) \C_u = {\mathbf P}_C 
\{\delta \H|_{e^{-i\omega t}}\} \C, \nonumber \\
& & \delta \H|_{e^{-i\omega t}} = 
\sum_{j=1}^{Q} \sum_{n_j,m_j=1}^{M_j} 
[(\{{h^j}^\ast\}_{m_j^\ast v_{n_j}} 
+ h^j_{n_j u_{m_j}} + \{f^\dag_j\}_{n_jm_j})\brho_{n_jm_j}^j + \nonumber \\
& & \qquad  + \sum_{\vec n[j],\vec m[j]} 
(W_{\vec n[j]\vec m[j]m_j^\ast v_{n_j}} + 
 W_{\vec n[j]\vec m[j]n_ju_{m_j}})\brho_{\vec n\vec m}] + 
\sum_{\vec n,\vec m} \{g^\dag\}_{\vec n\vec m} \brho_{\vec n\vec m}. \
\eeqn
Similarly,
the first-order equation associated with the $e^{+i\omega t}$ term reads:
\beqn\label{coeff_LR_MCTDH_3}
& & {\mathbf P}^\ast_C (\varepsilon - \omega - \H^\ast) \C_v = 
{\mathbf P}^\ast_C \{\delta \H|_{e^{+i\omega t}}\}^\ast \C^\ast, \nonumber \\
& & \{\delta \H|_{e^{+i\omega t}}\}^\ast = 
\sum_{j=1}^{Q} \sum_{n_j,m_j=1}^{M_j} 
[(h^j_{m_j u_{n_j}} 
+ \{{h^j}^\ast\}_{n^\ast_j v_{m_j}} + \{f^\dag_j\}_{m_jn_j})\brho_{n_jm_j}^j + \nonumber \\
& & \qquad  + \sum_{\vec n[j],\vec m[j]} 
(W_{\vec m[j]\vec n[j]m_j u_{n_j}} + 
 W_{\vec m[j]\vec n[j]n^\ast_jv_{m_j}})\brho_{\vec n\vec m}] + 
\sum_{\vec n,\vec m} \{g^\dag\}_{\vec m\vec n} \brho_{\vec n\vec m}. \
\eeqn
Compare Eqs.~(\ref{coeff_LR_MCTDH_2}) and (\ref{coeff_LR_MCTDH_3}) to the identical-particle case,
Eqs.~(\ref{LR_1st_coeff_2}) and (\ref{LR_1st_coeff_3}), respectively.

%%%%%%%%%%%%%%%%%%%%%%%%%%%%%%%%%%%%%%%%%%%%%%%%
\subsubsection{Casting the linear-response equations into a matrix form}\label{MCTDH_sub_LR_eqs_s2}
%%%%%%%%%%%%%%%%%%%%%%%%%%%%%%%%%%%%%%%%%%%%%%%%

The quantum object made of the $Q$ coupled degrees-of-freedom responds 
to the external perturbation as a whole.
We thus combine the response amplitudes of all orbitals and expansion coefficients 
of a given perturbed wavefunction together.
The combined response vector: 
\beq\label{LR_MCTDH_vector}
\begin{pmatrix} 
 \u^1 \\
 \vdots \\
 \u^Q \\
 \v^1 \\
 \vdots \\
 \v^Q \\
 \C_u \\
 \C_v \\
\end{pmatrix},
\qquad \qquad 
\u^j = \{|u_{m_j}\rangle\}, \
\v^j = \{|v_{m_j}\rangle\}, \ \qquad
j=1,\ldots,Q,\ 
m_j=1,\ldots,M_j, 
\eeq
is now of length $2(\sum_{j=1}^{Q}M_j+N_{\mathit{conf}})$.

Following the strategy taken for identical particles,
the {\it final} 
result for the linear-response working matrix equation in the orbital--coefficient response space is given by:
\beq\label{LR_matrix_EO_working_mctdh}
\left(\bcalL-\omega\right)
\begin{pmatrix} 
 \u^1 \\
 \vdots \\
 \u^Q \\
 \v^1 \\
 \vdots \\
 \v^Q \\
 \C_u \\
 \C_v \\
\end{pmatrix} =
\bcalR.
\eeq
The linear-response matrix $\bcalL$ and the vector $\bcalR$ which collects the various perturbing fields,
see Eq.~(\ref{MCTDH_LR_ansatz}),
are constructed explicitly below.
Analogously to Eq.~(\ref{LR_matrix_EO_working}),
we term Eq.~(\ref{LR_matrix_EO_working_mctdh}) LR-MCTDH theory.

The solution of the LR-MCTDH linear-response matrix system, Eq.~(\ref{LR_matrix_EO_working_mctdh}),
and of the distinguishable-particle Schr\"odinger equation in linear response
requires the 
eigenvalues $\{\omega_k\}$ and eigenvectors $\{\R^k\}$
of the linear-response matrix $\bcalL$: 
\beq\label{LR_matrix_eigenvalues_prob_mctdh}
\bcalL
\begin{pmatrix} 
 \u^{1,k} \\
 \vdots \\
 \u^{Q,k} \\
 \v^{1,k} \\
 \vdots \\
 \v^{Q,k} \\
 \C_u^k \\
 \C_v^k \\
\end{pmatrix} =
\omega_k
\begin{pmatrix} 
 \u^{1,k} \\
 \vdots \\
 \u^{Q,k} \\
 \v^{1,k} \\
 \vdots \\
 \v^{Q,k} \\
 \C_u^k \\
 \C_v^k \\
\end{pmatrix} \equiv \omega_k \R^k.
\eeq
This task is accomplished in Subsec.~\ref{MCTDH_sub_LR_eqs_s3} below.

Now, the matrix form of the linear-response 
equations of LR-MCTDH, Eq.~(\ref{LR_matrix_EO_working_mctdh}), is assembled as follows.
Just like in the identical-particle case, Eq.~(\ref{LR_matrix}),
the linear-response matrix $\bcalL$ is divided into 4 blocks,
$\bcalL_{oo}$, $\bcalL_{oc}$, $\bcalL_{co}$, and $\bcalL_{cc}$.
The orbital--orbital block $\bcalL_{oo}$
is further divided into four 
sub-matrices, like Eq.~(\ref{LR_matrix_oo}),
with the $\bcalL_{oo}^u$ and $\bcalL_{oo}^v$
sub-matrices are additionally divided into 
$Q \times Q$ rectangular (square on the diagonal) 
sub-parts each of dimension $M_j \times M_k$ 
as follows
($j,k=1,\ldots,Q$):
\beqn\label{LR_MCTDH_matrix_oo_1}
\bcalL_{oo}^{u,jj} = & & \!\!\! \brho^j \hat h^j - \bmu^j + \bOmega^j, \nonumber \\
& & \!\!\! \brho^j = \{\rho^j_{n_jm_j}\}, \ 
    \bmu^j = \{\mu^j_{n_jm_j}\}, \
    \bOmega^j = \left\{\hat \Omega^j_{n_jm_j}\right\}, \nonumber \\
\bcalL_{oo}^{u,jk} = & & \!\!\! \bkappa^{1,jk} = \{\kappa^{1,jk}_{n_jm_k}\} =
\left\{ \sum_{\vec n[j],\vec m[k]} 
\rho_{\vec n \vec m} \hat K_{\vec n[j]\vec m[k]} \right\}, 
\qquad k\ne j, \nonumber \\
\bcalL_{oo}^{v,jj} = & & \!\!\! {\0 }_j, \nonumber \\
\bcalL_{oo}^{v,jk} = & & \!\!\! \bkappa^{2,jk} = \{\kappa^{2,jk}_{n_jm_k}\} =
\left\{ \sum_{\vec n[j],\vec m[k]} 
\rho_{\vec n \vec m} \hat K_{\vec n[j,k] \vec m} \right\},
\qquad k\ne j,  \
\eeqn
where ${\0 }_j$ is a unit matrix of dimension $M_j$.
Note that the diagonal sub-parts $\bcalL_{oo}^{v,jj}$ of $\bcalL_{oo}^v$ are zero.
This should be contrasted with the respective situation
for identical particles,
see Eq.~(\ref{LR_matrix_oo_1}),
where the diagonal sub-matrix $\bcalL_{oo}^v$ is non-zero.
The reason is distinguishability, 
namely,
that in MCTDH there are no two particles associated with the {\it same} degree-of-freedom.
Consequently,
there are no exchange terms within the same degree-of-freedom 
of the LR-MCTDH linear-response matrix,
compare Eqs.~(\ref{LR_matrix_oo_1}) and (\ref{LR_MCTDH_matrix_oo_1}).
As can be seen in the latter, 
there are no exchange operators in neither
the $\bcalL_{oo}^{u,jj}$ nor the $\bcalL_{oo}^{v,jj}$ diagonal sub-parts.

The orbital--coefficient block $\bcalL_{oc}$, like Eq.~(\ref{LR_matrix_oc}),
is comprised of the four sub-matrices where
($j=1,\ldots,Q, n_j=1,\ldots,M_j$):
\beqn\label{LR_MCTDH_matrix_oc_1}
 & & \bcalL_{oc}^{u,j} = 
\left\{ \C^\dag \sum_{m_j=1}^{M_j}
[\hat h^j  \brho_{n_jm_j}^j 
+ \sum_{\vec n[j],\vec m[j]} \hat W_{\vec n[j] \vec m[j]}
 \brho_{\vec n \vec m}]
|m_j\rangle
\right\},
\nonumber \\
 & & \bcalL_{oc}^{v,j} =  
\left\{\C^t \sum_{m_j=1}^{M_j}
 [\hat h^j \brho_{m_jn_j}^j
+ \sum_{\vec n[j],\vec m[j]} \hat W_{\vec n[j] \vec m[j]} \brho_{\vec m \vec n}] 
|m_j\rangle 
\right\}. \
\eeqn
The coefficient--orbital block, like Eq.~(\ref{LR_matrix_co}),
is comprised of the four sub-matrices with
($j=1,\ldots,Q, n_j=1,\ldots,M_j$):
\beqn\label{LR_MCTDH_matrix_co_1}
 & & \bcalL_{co}^{u,j} = 
\left\{ \sum_{m_j=1}^{M_j} \langle m_j| 
[\hat h^j  \brho_{m_jn_j}^j 
+ \sum_{\vec n[j],\vec m[j]} \hat W_{\vec m[j] \vec n[j]}
 \brho_{\vec m \vec n}] \C
\right\},
\nonumber \\
 & & \bcalL_{co}^{v,j} =  
\left\{\sum_{m_j=1}^{M_j} \langle m_j^\ast | 
 [ \{{h^j}^\ast\} \brho_{n_jm_j}^j
+ \sum_{\vec n[j],\vec m[j]} \hat W_{\vec n[j] \vec m[j]} \brho_{\vec n \vec m}] \C
\right\}. \
\eeqn
Furthermore, 
the relation between the two off-diagonal rectangular blocks 
of $\bcalL$ as in Eq.~(\ref{LR_matrix_oc_co_relation}),
namely,
$(\bcalL_{oc}^{u,j})^\dag = (\bcalL_{co}^{u,j})$ and 
$(\bcalL_{oc}^{v,j})^{t} = (\bcalL_{co}^{v,j})$, $j=1,\ldots,Q$,
holds.
Finally, the coefficients--coefficients block is given, 
like Eq.~(\ref{LR_matrix_cc_mapping}), as
\footnote{There is no difference between the $\star$ atop the
identical-particle second-quantized Hamiltonian in Eqs.~(\ref{LR_1st_coeff_3}) and (\ref{LR_matrix_cc_mapping})
and the $\ast$ atop the distinguishable-degrees-of-freedom Hamiltonian-matrix in
Eq.~(\ref{LR_MCTDH_matrix_cc}),
because the density-operators' matrices, see Appendix \ref{MCTDH_tensor_appendix}, are real-valued matrices.}: 
\beq\label{LR_MCTDH_matrix_cc}
 \bcalL_{cc} =
\begin{pmatrix} 
 \H - \varepsilon {\1 }_c & {\0 }_c \\
  {\0 }_c & -(\H^\ast - \varepsilon {\1 }_c) \\
\end{pmatrix},
\eeq
where $\H$ is given in Eq.~(\ref{n_Ham}) and, to recall,
${\1 }_c$ and ${\0 }_c$ are 
the unit and zero matrices
of dimension $N_{\mathit{conf}}$.

We proceed with the ingredients needed for the linear-response matrix system,
as done in Subsec.~\ref{matrix_form_sec}.
The combined orbitals--coefficients projector reads:
\beq\label{LR_matrix_P_mctdh}
 \bcalP =
\begin{pmatrix} 
 \bcalP_o & {\0 }_{oc} \\
  {\0 }_{co} &  \bcalP_c \\
\end{pmatrix}, \qquad \quad
 \bcalP_o =
\begin{pmatrix} 
 {\mathbf P} & {\0 }_o \\
  {\0 }_o &  {\mathbf P}^\ast \\
\end{pmatrix}, \
{\mathbf P} = 
\begin{pmatrix} 
 \hat{\mathbf P}_1 {\1 }_o^1 & \cdots & {\0 }_o^{1Q} \\
 \vdots  & \ddots & \vdots  \\
 {\0 }_o^{Q1} & \cdots &  \hat{\mathbf P}_Q {\1 }_o^Q \\
 \end{pmatrix},
\ \quad
 \bcalP_c =
\begin{pmatrix} 
 {\mathbf P}_C & {\0 }_c \\
  {\0 }_c & {\mathbf P}^\ast_C \\
\end{pmatrix},
\eeq
where ${\mathbf P}$ collects on its diagonal the projectors $\{\hat{\mathbf P}_j\}, j=1,\ldots,Q$ 
of all degrees-of-freedom,
and the dimension of the various unit and zero 
matrices appearing in Eq.~(\ref{LR_matrix_P_mctdh}) is obvious.
Similarly,
the combined orbital--coefficient metric reads:
\beq\label{LR_matrix_M_mctdh}
 \bcalM =
\begin{pmatrix} 
 \brho_o & {\0 }_{oc} \\
  {\0 }_{co} &  {\1 }_{2c} \\
\end{pmatrix}, \qquad \qquad
 \brho_o =
\begin{pmatrix} 
 \brho  & {\0 }_o \\
  {\0 }_o &   \brho^\ast \\
\end{pmatrix}, \
 \brho = 
\begin{pmatrix} 
  \brho^1 & \cdots & {\0 }_o^{1Q} \\
 \vdots  & \ddots & \vdots  \\
 {\0 }_o^{Q1} & \cdots & \brho^Q \\
 \end{pmatrix},
\eeq
where $\brho$ collects on its diagonal the reduced one-body density matrices $\{\brho^j\}, j=1,\ldots,Q$ 
of all degrees-of-freedom.

To continue, let us collect the perturbing fields in the vector:
\beq\label{LR_perturbing_fields_mctdh}
 \bcalM^{+\frac{1}{2}} \bcalR = \bcalM^{+\frac{1}{2}} 
\begin{pmatrix} 
 - \hat f_1^\dag \bphi_1 \\
 \vdots \\
 - \hat f_Q^\dag \bphi_Q \\
   \hat f_1^\ast \bphi_1^\ast \\
 \vdots \\
   \hat f_Q^\ast \bphi_Q^\ast \\
 - \sum_{j=1}^{Q} \sum_{n_j,m_j=1}^{M_j} \{f^\dag_j\}_{n_jm_j}\brho_{n_jm_j}^j \C \\
   \sum_{j=1}^{Q} \sum_{n_j,m_j=1}^{M_j}  \{f^\dag_j\}_{m_jn_j}\brho_{n_jm_j}^j \C^\ast \\
\end{pmatrix} + 
\bcalM^{-\frac{1}{2}} 
\begin{pmatrix} 
 - \bOmega^1_g \bphi_1 \\
 \vdots \\
 - \bOmega^Q_g \bphi_Q \\
   {\bOmega^1_g}^\ast \bphi_1^\ast \\
 \vdots \\
  {\bOmega^Q_g}^\ast \bphi_Q^\ast \\
 - \sum_{\vec n,\vec m} \{g^\dag\}_{\vec n\vec m} \brho_{\vec n\vec m} \C \\
 \sum_{\vec n,\vec m} \{g^\dag\}_{\vec m\vec n} \brho_{\vec n\vec m} \C^\ast \\
\end{pmatrix},
\eeq
where $\bphi_j=\{|\phi_{m_j}\rangle\},\, j=1,\ldots,Q,\, m_j=1,\ldots, M_j$ are column vectors
and
$\bOmega^j_g = \{\Omega^j_{g,n_jm_j}\} 
= \left\{\sum_{\vec n[j],\vec m[j]} \rho_{\vec n \vec m} 
    \{\hat g^\dag\}_{\vec n[j] \vec m[j]}\right\}$
are square matrices of dimensions $M_j \times M_j,\, j=1,\ldots,Q$.
Just as is done in Subsec.~\ref{matrix_form_sec} utilizing the 
assignments 
$\left\{\bcalP \bcalM^{-\frac{1}{2}}\bcalL\bcalM^{-\frac{1}{2}}\bcalP\right\}
\Longrightarrow \bcalL$, 
$\left\{\bcalP \bcalM^{+\frac{1}{2}}\bcalR\right\} \Longrightarrow \bcalR$, 
and
$\left\{\bcalM^{+\frac{1}{2}} 
\begin{pmatrix} 
 \u^1 \ \ldots \ \u^Q \ \v^1 \ \ldots \ \v^Q \ \C_u \ \C_v
\end{pmatrix}^t\right\} \Longrightarrow 
\begin{pmatrix} 
 \u^1 \ \ldots \ \u^Q \ \v^1 \ \ldots \ \v^Q \ \C_u \ \C_v
\end{pmatrix}$,
we arrive at the linear-response matrix equation (\ref{LR_matrix_EO_working_mctdh}).

%%%%%%%%%%%%%%%%%%%%%%%%%%%%%%%%%%%%%%%%%%%%%%%%%
\subsubsection{Solving the distinguishable-particle Schr\"odinger equation in linear response}\label{MCTDH_sub_LR_eqs_s3}
%%%%%%%%%%%%%%%%%%%%%%%%%%%%%%%%%%%%%%%%%%%%%%%%%

To solve the LR-MCTDH linear-response system, Eq.~(\ref{LR_matrix_EO_working_mctdh}),
we have to diagonalize the linear-response matrix $\bcalL$
and find its
excitations energies $\{\omega_k\}$ and eigenvectors $\{\R^k\}$. 
The analysis of the symmetries and subsequent eigenstates' resolutions
follow exactly the route 
of the identical-particle linear-response LR-MCTDHX.
We capture the essence here for completeness.

We begin with the `spin' matrices 
${\bSig}_\1 =\begin{pmatrix} 
 {\bSig}_{\1 }^o & {\0 }_{oc} \\
 {\0 }_{co} & {\bSig}_{\1 }^c \\
\end{pmatrix}$
and
${\bSig}_\3 =
\begin{pmatrix} 
 {\bSig}_{\3 }^o & {\0 }_{oc} \\
 {\0 }_{co} & {\bSig}_{\3 }^c \\
\end{pmatrix}$.
They have exactly the same appearance and structure as in the identical-particle system, 
Eqs.~(\ref{LR_matrix_Sig1}) and (\ref{LR_matrix_Sig3}), 
the only difference is the dimension of their orbital
entires being $\sum_{j=1}^{Q}M_j$. 
Now, let $\R^k$ 
be the right eigenvector of $\bcalL$ with the (real) eigenvalue $\omega_k$.
From the symmetries ${\bSig}_\3 \bcalL {\bSig}_\3 = (\bcalL)^\dag$ and 
${\bSig}_\1 \bcalL {\bSig}_\1 = -(\bcalL)^\ast$
we find that $\L^k = \mathrm{sng}^k {\bSig}_\3 \R^k$
[where $\mathrm{sng}^k$ stands for the sign of the `scalar product' $\{(\R^k)^\dag {\bSig}_\3 \R^k\}$],
that $\R^{-k} \equiv {\bSig}_\1 (\R^k)^\ast$ and $\L^{-k} = - \mathrm{sng}^k {\bSig}_\3 \R^{-k}$,
and the orthogonality relations
$(\L^k)^\dag \R^{k'} = \mathrm{sng}^k \left[\sum_{j=1}^Q 
\left\{(\u^{j,k})^\dag \u^{j,k'} - (\v^{j,k})^\dag \v^{j,k'} \right\} 
+ (\C_u^k)^\dag \C_u^{k'} - (\C_v^k)^\dag \C_v^{k'}
\right] = \delta_{kk'}$
and 
$(\L^k)^\dag \R^{-k'} = \mathrm{sng}^k \left[\sum_{j=1}^Q 
\left\{(\u^{j,k})^\dag (\v^{j,k'})^\ast - (\v^{j,k})^\dag (\u^{j,k'})^\ast \right\} 
+ (\C_u^k)^\dag (\C_v^{k'})^\ast - (\C_v^k)^\dag (\C_u^{k'})^\ast
\right] = 0$.
From these, the resolutions of the identify and of the response matrix follow:
\beqn\label{LR_resolutions_mctdh}
& & \1 = \sum_{k=0}^{N_0}  \left\{ \R_0^{k} (\L_0^{k})^\dag + \R_0^{-k} (\L_0^{-k})^\dag \right\}
+ \sum_{k>N_0} \left\{ \R^{k} (\L^{k})^\dag + \R^{-k} (\L^{-k})^\dag \right\}, 
\qquad
N_0 = \sum_{j=1}^Q M_j^2,
\nonumber \\
& & \bcalL = \sum_{k>N_0} \omega_k \left\{ \R^{k} (\L^{k})^\dag -
              \R^{-k} (\L^{-k})^\dag \right\}. \
\eeqn
The first group of vectors in Eq.~(\ref{LR_resolutions_mctdh}) are the zero-mode excitations.
Comparison of Eq.~(\ref{LR_resolutions_mctdh}) for distinguishable degrees-of-freedom to the 
identical-particle case, Eqs.~(\ref{LR_identity_resolution}) and (\ref{LR_calL_matrix_resolution}),
shows that the number of zero modes $N_0$ is the sum of the squares of $\{M_j\}, j=1,\dots,Q$,
i.e., the number of orbitals used in the ground-state wavefunction for each degree-of-freedom.
The second group of vectors are the non-zero-mode excitations.
Like above, their summation index, $k>N_0$,
indicates that we enlist them after the group of zero-mode excitations
for which the index of enumeration satisfies $k \in [0,N_0]$.

Next, expanding the response amplitudes and perturbation with the eigenvectors of
$\bcalL$ and substituting into Eq.~(\ref{LR_matrix_EO_working_mctdh}), 
we obtain the explicit expression for the system's response vector and response weights:
\beqn\label{LR_expand_mctdh}
& & \qquad \qquad 
\begin{pmatrix} 
 \u^1 \\
 \vdots \\
 \u^Q \\
 \v^1 \\
 \vdots \\
 \v^Q \\
 \C_u \\
 \C_v \\
\end{pmatrix} = \sum_{k>N_0} \left[\frac{\gamma_k}{\omega-\omega_k} \R^{k} 
+ \frac{\gamma_{-k}}{\omega+\omega_{k}} \R^{-k}\right], \nonumber \\
& &  \gamma_k = - (\L^{k})^\dag \bcalR = \mathrm{sng}^k \Bigg \{ \sum_{j=1}^Q \Big( 
\{\u^{j,k}\}^\dag \left[\{\brho^j\}^{+\frac{1}{2}} \hat f^\dag_j 
+ \{\brho^j\}^{-\frac{1}{2}}\bOmega^j_g\right] \bphi_j + \nonumber \\
& & \qquad + \{\v^{j,k}\}^\dag  \left[\{\brho^j\}^{+\frac{1}{2}} \hat f^\ast_j 
+ \{\brho^j\}^{-\frac{1}{2}} {\bOmega^j_g}^\ast\right] \bphi_j^\ast \Big)+ \nonumber \\
& & \qquad + (\C_u^k)^\dag \cdot 
\left[\sum_{j=1}^{Q} \sum_{n_j,m_j=1}^{M_j} \{f^\dag_j\}_{n_jm_j}\brho_{n_jm_j}^j 
+ \sum_{\vec n,\vec m} \{g^\dag\}_{\vec n\vec m} \brho_{\vec n\vec m} \right] \C + \nonumber \\
& &   \qquad   +  (\C_v^k)^\dag \cdot
\left[\sum_{j=1}^{Q} \sum_{n_j,m_j=1}^{M_j}  \{f^\dag_j\}_{m_jn_j}\brho_{n_jm_j}^j +
\sum_{\vec n,\vec m} \{g^\dag\}_{\vec m\vec n} \brho_{\vec n\vec m} \right]\C^\ast, \nonumber \\
& &  \gamma_{-k} = - (\L^{-k})^\dag \bcalR = 
- \mathrm{sng}^k \Bigg \{ \sum_{j=1}^Q \Big( 
\{\v^{j,k}\}^t \left[ \{\brho^j\}^{+\frac{1}{2}} \hat f^\dag_j + \{\brho^j\}^{-\frac{1}{2}}\bOmega_g^j\right] \bphi_j + \nonumber \\
& & \qquad +  \{\u^{j,k}\}^t \left[ \{\brho^j\}^{+\frac{1}{2}} \hat f^\ast_j 
+ \{\brho^j\}^{-\frac{1}{2}} {\bOmega^j_g}^\ast\right] \bphi^\ast_j \Big) + \nonumber \\
& & \qquad + (\C_v^k)^t \cdot 
\left[\sum_{j=1}^{Q} \sum_{n_j,m_j=1}^{M_j} \{f^\dag_j\}_{n_jm_j}\brho_{n_jm_j}^j 
+ \sum_{\vec n,\vec m} \{g^\dag\}_{\vec n\vec m} \brho_{\vec n\vec m} \right] \C + \nonumber \\
& &   \qquad   +  (\C_u^k)^t \cdot 
\left[\sum_{j=1}^{Q} \sum_{n_j,m_j=1}^{M_j}  \{f^\dag_j\}_{m_jn_j}\brho_{n_jm_j}^j +
\sum_{\vec n,\vec m} \{g^\dag\}_{\vec m\vec n} \brho_{\vec n\vec m} \right]\C^\ast. \
\eeqn
Note that the response weighs $\{\gamma_{k},\gamma_{-k}\}$
incorporate the response of the {\it entire} system of the $Q$ coupled degrees-of-freedom,
via their orbitals and expansion coefficients. 

From the solution to the response amplitudes, Eq.~(\ref{LR_expand_mctdh}), 
and the ansatz for the orbitals and expansion coefficients, Eq.~(\ref{MCTDH_LR_ansatz}), 
we get the 
time dependence of the latter in linear response:
\beqn\label{orb_coeff_LR_explicit}
& & \bphi_j(\r_j,t) \approx \bphi_j(\r_j) + \delta\bphi_j(\r_j,t), \qquad \qquad j=1,\ldots,Q, \nonumber \\
& & \qquad \delta\bphi_j(\r,t) = \sum_{k>N_0} 
 \Bigg\{ \left[\gamma_k  \{\brho^j\}^{-\frac{1}{2}} \u^{j,k}(\r_j) e^{-i \omega t} + 
       \gamma_k^\ast \{{\brho^j}^\ast\}^{-\frac{1}{2}}  \{\v^{j,k}(\r_j)\}^\ast e^{+i \omega t} \right]/(\omega-\omega_k) + \nonumber \\
& & \qquad \qquad + \left[\gamma_{-k}  \{\brho^j\}^{-\frac{1}{2}}  \{\v^{j,k}(\r_j)\}^\ast e^{-i \omega t} + 
       \gamma_{-k}^\ast \{{\brho^j}^\ast\}^{-\frac{1}{2}}  \u^{j,k}(\r_j) e^{+i \omega t} \right]/(\omega+\omega_k) \Bigg\}, \nonumber \\
& & \C(t) \approx \C + \delta\C(t), \nonumber \\
& & \qquad \delta\C(t) = \sum_{k>N_0} 
\Bigg\{ \left[\gamma_k \C_u^k e^{-i \omega t} + 
       \gamma_k^\ast  \{\C_v^k\}^\ast e^{+i \omega t} \right]/(\omega-\omega_k) + \nonumber \\
& & \qquad \qquad + \left[\gamma_{-k}  \{\C_v^k\}^\ast  e^{-i \omega t} + 
       \gamma_{-k}^\ast  \C_u^k e^{+i \omega t} \right]/(\omega+\omega_k) \Bigg\},
\eeqn
with $\delta\bphi_j(\r,t) = \{\delta\phi_{m_j}(\r_j,t)\}, m_j=1,\ldots,M_j$.
As before,
the orbitals and the expansion coefficients show the largest
response at the frequencies $\{\pm \omega_k\}$. 
From Eq.~(\ref{orb_coeff_LR_explicit}) and to complete out task,
the time-dependent many-particle wavefunction is given in 
linear response by:
\beqn\label{LR_wavefunction_mctdh}
& & |\Psi(t)\rangle \approx \sum_{\vec n} C_{\vec n} |\vec n\rangle +   \sum_{\vec n} \delta C_{\vec n}(t) |\vec n\rangle +  \\
& & \quad + 
\sum_{\vec n} C_{\vec n}
 \left[ \sum_{j=1}^{Q} \sqrt{\langle \delta\phi_{n_j}(\r,t)|\delta\phi_{n_j}(\r,t)\rangle}
   |\vec n[j], 1_{M_j+1}^{n_j};t\rangle\right], \nonumber \
\eeqn
where $|1_{M_j+1}^{n_j};t\rangle$ is associated with the (unnormalized) time-dependent $n_j$-th response orbital $\delta\phi_{n_j}(\r_j,t)$
of the $j$-th degrees-of-freedom.
Similarly to above, 
the response orbitals $\{\delta\phi_{n_j}(\r_j,t)\}$ need not be orthogonal to each other,
unlike their orthogonality with the ground-state orbitals of the $j$-th degree-of-freedom which originates from
the orbital differential condition (\ref{diff_cond_MCTDH_both}).
This brings our derivation of LR-MCTDH to a completion.

%%%%%%%%%%%%%%%%%%%%%%%%%%%%%%%%%%%%%%%%%%%%%%%%
\section{Summary and Concluding Remarks}\label{sum_con_sec}
%%%%%%%%%%%%%%%%%%%%%%%%%%%%%%%%%%%%%%%%%%%%%%%%

In this work we present  
a unified representation and 
view on linear response of
systems of interacting particles, 
let them be identical or distinguishable.
The approach is 
based on the numerically-exact equations of motion of various 
multiconfigurational time-dependent Hartree (MCTDH) theories.
Linearizing MCTDHB and MCTDHF leads to 
the linear-response theories
LR-MCTDHB and LR-MCTDHF for identical bosons
and fermions, respectively,
whereas linearizing MCTDH for distinguishable degrees-of-freedom leads
to the respective theory LR-MCTDH.
Thus, these linear-response theories provide 
numerically-exact
excitation energies and  also system's properties, 
via the response operators and amplitudes,
when numerical convergence is achieved in the calculations.

As a complementary result 
we introduce an additional projector operator onto the 
coefficients' part of MCTDH methods, which leads
to a new differential condition for the expansion coefficients within these methods.
Together with the famous (orbital) differential condition \cite{MCTDH_cpl,MCTDH_jcp},
these two lead to the notion of ``fully-projected'' MCTDH, and MCTDHB and MCTDHF equations of motion,
namely, that the time-dependent multiconfigurational wavepackets 
evolve in the completely orthogonal space.
As a result, 
the present linear-response
theories ensure the response of the orbitals and expansion coefficients to be orthogonal 
to the
stationary (ground-state) wavefunction, 
because of the appearing orbitals' and 
coefficients' projector operators.

We analyze in particular and explicitly
one-body and two-body response
operators for identical particles and up to 
all-system response operators for distinguishable
degrees-of-freedom.
The resulting working linear-response equations
are presented and discussed in detail.
The response matrix, which provides the desired excitation
energies, 
does not depend on the special form of the perturbing fields. 
Consequently, 
the choice of the perturbing fields can be utilized to study
the nature of the respective excited states.
Generally, higher-body response operators can give access to
more involved excitations.
In particular for Bose-Einstein condensates,
where experimentally it is a standard practice to alter
the particle-particle (two-body) interactions 
\cite{Book_Pitaevskii,Book_Pethick}, 
one could naturally expect to access and subsequently analyze 
new classes of 
many-body excitations. 

The steps to be followed in order to solve the many-body problem in linear response 
-- for identical or distinguishable interacting particles --
may be summarized by the following flowchart.
First, we calculate the ground state.
This is done at a certain level of MCTDHB, MCTDHF, or MCTDH.
Second, we construct the linear-response matrix $\bcalL$.
Third, we diagonalize $\bcalL$ to obtain its eigenvalues -- the excitations
energies $\{\omega_k\}$ -- and eigenvectors $\{\R^k\}$.
Fourth, for a particular choice of the perturbing fields, collected in the response vector $\bcalR$, 
the eigenvectors are utilized 
to compute the response weights $\{\gamma_k\}$, 
which quantify the intensity of the response.
Fifth, all these ingredients are combined together to prescribe the
time-dependent orbitals and expansion coefficients and hence
the many-particle wavefunction $|\Psi(t)\rangle$ in linear response.
The computation of observables and system's properties within
the linear-response theories presented here,
LR-MCTDHB, LR-MCTDHF, and LR-MCTDH, has not been discussed explicitly 
in the present work, and is deferred to elsewhere.

It is possible to compute low-lying excited states directly by MCTDH methods, 
see, e.g., Ref.~\cite{MCTDH_book}.
The resulting wavefunctions are stationary solutions and hence
can in principle be utilized as inputs in the above-summarized linear-response
procedure. The resulting spectrum and response amplitudes would
describe the respective excitations and de-excitations atop these states. 

Until now we discussed linear-response theory in bound-state systems.
The same formalism can in principle be extended to metastable states. 
By analytically continuing the response matrix into the complex energy plane, 
one can also compute metastable excited states of the system and their lifetimes. 
There are several methods available in the literature to carry out the analytic continuation. 
One is complex rotation, see, e.g., the reviews \cite{NIM_review,NIM_book}, 
and another is by adding a complex absorbing potential to the Hamiltonian of the system,
see, e.g., the review \cite{RL_review}. 
Extending the theory for the linear-response approaches to include 
metastable states might be more involved than for the reported ones 
and is not available yet.

The ideas presented here can also be extended to other
derived theories emanating from the MCTDH theory,
such that for mixtures of identical particles 
\cite{MCTDH_MIX,3MIX},
also see Refs.~\cite{Kato_Yamanouchi1,Kato_Yamanouchi2}.
The separation of the coefficients, which define the reduced density matrices,
from the orbitals \cite{MCHB}
suggests that other representations for the time-dependent many-body wavepackets
would be amenable to linear response in the spirit presented here.
For instance, it is possible to envision that
propagation theories utilizing other multiconfigurational ans\"atze -- such as, e.g., 
in multilayer-formulated MCTDH methods 
\cite{ML_original,ML_Manthe,ML_Oriol_Dieter,ML_MCTDHB} --
would lead to appealing linear-response theories as well.  
Recent fruitful implementation and applications for trapped identical bosons \cite{LR_MCTDHB}
suggest that much more is to be expected from MCTDH-based linear-response theories.

\acknowledgments
We are grateful to Julian Grond for multiple fruitful discussions,
as well as to Axel Lode and Kaspar Sakmann.
Financial support by the DFG is acknowledged.
OEA is grateful to the continuous hospitality
of the Lewiner Institute for Theoretical Physics (LITP)
at the Department of Physics, Technion.

\appendix

%%%%%%%%%%%%%%%%%%%%%%%%%%%%%%%%%%%%%%%%
\section{The differential condition in MCTDHB and MCTDHF}\label{dif_con_proof_appendix}
%%%%%%%%%%%%%%%%%%%%%%%%%%%%%%%%%%%%%%%%

The differential condition Eq.~(\ref{diff_cond}) introduced by the MCTDH 
developers 
into the field of multi-dimensional wavepacket propagation
\cite{MCTDH_cpl,MCTDH_jcp} is what has made the MCTDH and its 
daughters such efficient, effective, and practical propagation theories and tools.
We have seen that in the context of 
linear-response theories the differential condition for the orbitals, Eq.~(\ref{diff_cond}), 
together with its analog differential condition for the coefficients, Eq.~(\ref{C_diff_cond}),
are what enables us a fully orthogonal response space [also see Eq.~(\ref{diff_cond_MCTDH_both})].
It is thus 
instrumental that we recall in this appendix
how exactly the differential condition Eq.~(\ref{diff_cond}) can 
be introduced into the MCTDHX (X=B,F)
equations of motion, namely, what is the unitary transformation involved.   

The equations of motion for the orbitals and expansion coefficients as derived directly
from the MCTDHX action--functional are given by (see Refs.~\cite{MCTDHB2,unified}):
\beqn\label{pre_dif_MCTDHX_equ}
& &  \hat {\mathbf P} \sum_{q=1}^M [\rho_{pq} (\hat h - i\frac{\partial}{\partial t}) +
 \sum^M_{s,l=1}\rho_{pslq} \hat{W}_{sl}] |\phi_q\rangle = 0, 
\qquad p=1,\ldots,M, \nonumber \\
& & \bcalH (t)\C(t) = i\frac{\partial \C(t)}{\partial t}, \qquad
 {\mathcal H}_{\vec{n}\vec{n}'}(t) = \langle\vec{n};t|\hat H - i\frac{\partial}{\partial t}|\vec{n}';t\rangle.
\eeqn
We are looking for such a transformation on the orbitals and expansion coefficients
which leaves the many-particle wavefunction invariant, 
\beq\label{Psi_ALL_invariance}
 |\Psi(t)\rangle = \sum_{\vec n} C_{\vec n}(t) |\vec n;t\rangle = \sum_{\vec n} \overline{C}_{\vec n}(t) |\overline{\vec n;t}\rangle =
|\overline{\Psi}(t)\rangle,
\eeq
but removes the projector $\hat {\mathbf P}$ in front of the time derivative.
This is readily achieved by the unitary transformation:
\beq\label{unitary_phi2dot_1}
|{\overline{\phi}}_k\rangle = \sum_{p=1}^M |\phi_p\rangle U_{pk} \qquad \Longleftrightarrow \qquad
|\phi_q\rangle = \sum_{j=1}^M U^\ast_{qj} |{\overline{\phi}}_j\rangle,
\eeq 
with
\beq\label{unitary_phi2dot_2}
 i\dot U_{jq} = \sum_{k=1}^M -D_{jk} U_{kq}, \qquad D_{jk} = i\langle\phi_j|\dot\phi_k\rangle, \qquad j,k=1,\ldots,M,
\eeq
which leads explicitly to the differential condition Eq.~(\ref{diff_cond}) 
for the {\it transformed} orbitals:
\beq\label{unitary_phi2dot_3}
D_{\overline{j}\overline{k}}= i\langle\overline{\phi}_j|\dot{\overline{\phi}}_k\rangle=0, \qquad j,k=1,\ldots,M.
\eeq
In Eq.~(\ref{unitary_phi2dot_3}) 
we employ a shorthand notation where a 
transformed orbital ${\overline{\phi}}_k$
is denoted by $\overline{k}$.
Substituting Eq.~(\ref{unitary_phi2dot_1}) into the orbital part 
of Eq.~(\ref{pre_dif_MCTDHX_equ}),
making use of $\hat {\mathbf P} = \hat{\overline{\mathbf P}}$, 
of Eqs.~(\ref{Psi_ALL_invariance},\ref{unitary_phi2dot_3})
and
$i|\dot\phi_q\rangle = \sum_{j=1}^M (\dot U^\ast_{qj} | \overline{\phi}_j\rangle + U^\ast_{qj} |\dot{\overline{\phi}}_j\rangle)$,
$\hat{\overline{\mathbf P}} i|\dot\phi_q\rangle = \sum_{j=1}^M U^\ast_{qj} i|\dot{\overline{\phi}}_j\rangle$,
and that creation operators transform like 
Eq.~(\ref{unitary_phi2dot_1}),
one readily arrives 
at the relation:
\beq\label{pre_dif_MCTDHX_equ_2}
 \hat{\overline{\mathbf P}} \sum_{q=1}^M [\rho_{p\overline{q}}  \hat h +
 \sum^M_{s,l=1}\rho_{p\overline{s}\overline{l}\overline{q}} \hat{W}_{\overline{s}\overline{l}}] |\overline{\phi}_q\rangle = \sum_{q=1}^M \rho_{p\overline{q}} |\dot{\overline{\phi}}_q\rangle, 
\qquad p=1,\ldots,M. 
\eeq
Eq.~(\ref{pre_dif_MCTDHX_equ_2}) is expressed almost completely with
transformed orbitals, that is except of the leftmost creation operator denoted by $p$.
Next, multiplying both sides and summing up by $\sum_{p=1}^M U_{pk}$
(and removing the now superfluous `overline' from all quantities)
leads to the equations of motion for the orbitals of MCTDHX,
Eq.~(\ref{MCTDHX_equ}),
in their standard, unified
form \cite{unified,MCTDHB2}.

To show that the same simplification holds for the 
respective transformed coefficients and their equations of motion,
we do so by recalling that these equations are {\it form-invariant}
(see in this context Ref.~\cite{MCTDH_AM}).
Namely,
if
${\bcalH}(t)\C(t) = i\frac{\partial \C(t)}{\partial t}$
are satisfied for the untransformed quantities
$[\{C_{\vec{n}}(t)\}, \{\phi_k(\r,t)\}]$
then
$\overline{\bcalH}(t)\overline{\C}(t) = i\frac{\partial \overline{\C}(t)}{\partial t}$
are satisfied for the transformed ones
$[\{\overline{C}_{\vec{n}}(t)\}, \{\overline{\phi}_k(\r,t)\}]$.
The proof is straightforward.
Equating the variation
of the MCTDHX action--functional 
with respect to the expansion
coefficients to zero,
the result can be written as follows:
$\langle\vec{n};t|\hat H - i\frac{\partial}{\partial t}|\Psi(t)\rangle = 0, 
\forall \vec{n}$.
Since,
the transformed configurations $\{\langle\overline{\vec{n};t}|\}$
are given as
linear combinations of the untransformed configurations
$\{\langle\vec{n};t|\}$,
the operator $\hat H - i\frac{\partial}{\partial t}$
does not depend on the orbitals,
and making use of Eq.~(\ref{Psi_ALL_invariance})
we immediately get:
$\langle\overline{\vec{n};t}|
\hat H - i\frac{\partial}{\partial t}|\overline{\Psi}(t)\rangle = 0, \forall \vec{n}$,
which concludes our proof.
To our needs,
since the transformed orbitals Eq.~(\ref{unitary_phi2dot_1})
obey the differential condition Eq.~(\ref{unitary_phi2dot_3}),
the respective
equations of motion for
the transformed coefficients
(after removing the now superfluous `overline' from all quantities)
boil down to the standard equations of motion 
for the coefficients, Eq.~(\ref{MCTDHX_equ}).

%%%%%%%%%%%%%%%%%%%%%%%%%%%%%%%%%%%%%%%%
\section{Tensor-product representation of vectors 
and matrix elements in MCTDH}\label{MCTDH_tensor_appendix}
%%%%%%%%%%%%%%%%%%%%%%%%%%%%%%%%%%%%%%%%

In this appendix we present the tensor-product representation
of quantities in MCTDH, 
as far as they are needed in the derivation of LR-MCTDH.

The coefficients vector $\C$ in MCTDH can be written as follows:
\beq\label{T_E1}
 \C = \sum_{\vec n} C_{\vec n} \cdot {\1 }^1_{n_1} \otimes 
\cdots \otimes {\1 }^j_{n_j} \otimes \cdots \otimes {\1 }^Q_{n_Q},
\eeq
where ${\1 }^j_{n_j}$ is a column vector of zeros of length $M_j$,
except of $1$ in the $n_j$-th entry.
The time argument is suppressed throughout this appendix 
without loss
of generality. 
The density-operator matrices for the $j$-th degree-of-freedom can be written as,
$n_j, m_j = 1,\ldots,M_j$:
\beqn\label{T_E2}
& & \brho_{n_jm_j}^j = (\brho_{n_jm_j}^j)^\ast = 
{\1 }^{1} \otimes \cdots \otimes {\1 }^{j}_{n_jm_j} 
\otimes \cdots \otimes {\1 }^{Q}, \nonumber \\
& &  \brho_{m_jn_j}^j =
(\brho_{n_jm_j}^j)^t = {\1 }^{1} \otimes \cdots \otimes {\1 }^{j}_{m_jn_j} 
\otimes \cdots \otimes {\1 }^{Q}, \
\eeqn
where ${\1 }^{k}$ is a unit matrix of dimension $M_k$, 
and ${\1 }^{j}_{m_jn_j}$ is a square 
matrix of zeros of dimension $M_j$,
except of $1$ in the $(n_j,m_j)$-th entry.
From Eqs.~(\ref{T_E1}) and (\ref{T_E2}) we
get the elements of the reduced one-body density matrix Eq.~(\ref{MCTDH_r1}) 
-- written in a tensor-product form --
as expectation values 
using the coefficients' vector:
\beq\label{T_E3}
 \rho_{n_jm_j}^j = \C^\dag  \brho_{n_jm_j}^j \C. 
\eeq
We can now proceed in a recursive manner 
(see in this respect Ref.~\cite{3MIX}) and
assemble higher-order reduced densities.
From the two-degrees-of-freedom 
density operators
\beq\label{T_2B_dens_operator}
 \brho_{n_jn_km_jm_k}^{jk} = \brho_{n_jm_j}^j \brho_{n_km_k}^k =
{\1 }^{1} \otimes \cdots \otimes {\1 }^{j}_{n_jm_j} \otimes \cdots \otimes {\1 }^{k}_{n_km_k} \otimes \cdots \otimes {\1 }^{Q}, \qquad j\ne k
\eeq
the reduced two-degrees-of-freedom density matrix associated with
the $j$-th and $k$-the degrees-of-freedom is given by,
$n_j, m_j = 1,\ldots,M_j$ and $n_k, m_k = 1,\ldots,M_k$:
\beq\label{T_2B_dens_matrix}
\rho_{n_jn_km_jm_k}^{jk} = \C^\dag 
\brho_{n_jn_km_jm_k}^{jk} \C.
\eeq
This can be done with higher-order density-operators' matrices 
and reduced density matrices up to
the all-degrees-of-freedom ones:
\beq\label{T_AllB_dens_operator}
 \brho_{\vec n \vec m} = \brho_{n_1m_1}^1 \cdots \brho_{n_Qm_Q}^Q =
{\1 }^{1}_{n_1m_1} \otimes \cdots \otimes {\1 }^{Q}_{n_Qm_Q}
\eeq
and
\beq\label{T_AllB_dens_matrix}
 \rho_{\vec n \vec m} = \C^\dag \brho_{\vec n \vec m} \C 
= C^\ast_{\vec n} C_{\vec m}.
\eeq

From Eq.~(\ref{T_AllB_dens_matrix}) we 
get the elements of the mean-field operators Eq.~(\ref{MCTDH_MFs}) 
-- written in a tensor-product form -- as expectation values 
using the coefficients' vector:
\beq\label{T_E5}
 \hat \Omega^j_{n_jm_j} = 
 \sum_{\vec n[j],\vec m[j]} \hat W_{\vec n[j] \vec m[j]} \C^\dag \brho_{\vec n \vec m} \C,
\eeq
which enables us to readily perform the variation in Eq.~(\ref{LR_MCTDH_var_E1}).

Finally, the Hamiltonian in the basis of the configurations
can be written in an appealing tensor-product form.
To this end, consider the column vector of configurations:
\beq\label{n_1}
 \n^\dag 
= \sum_{\vec n} \langle\vec n| \cdot {\1 }^1_{n_1} \otimes 
\cdots \otimes {\1 }^j_{n_j} \otimes \cdots \otimes {\1 }^Q_{n_Q}.
\eeq
With this notation,
the matrix representation of the Hamiltonian Eq.~(\ref{Ham_MCTDH})
which is given in Eq.~(\ref{n_Ham}) 
takes on the form:
\beq\label{n_Ham_source}
\H = \n^\dag \hat H \m,
\eeq
which concludes our exposition.

%%%%%%%%%%%%%%%%%%%%%%%%%%%%%%%%%%%%%%%%%%%%%%%%

%%%%%%%%%%%%%%%%%%%%%%%%%%%%%%%%%%%%%%%%%%%%%%%%

\end{document}